\documentclass[acmsmall,screen,authorversion]{acmart}

\AtBeginDocument{%
  \providecommand\BibTeX{{%
    \normalfont B\kern-0.5em{\scshape i\kern-0.25em b}\kern-0.8em\TeX}}}

\setcopyright{acmlicensed}
\acmJournal{JRC}
\acmYear{2025} \acmVolume{1} \acmNumber{1} \acmArticle{1} \acmMonth{1}\acmDOI{10.1145/3764592}

\usepackage{tabularx}
\usepackage{booktabs}
\usepackage{multirow}
\usepackage[normalem]{ulem}
\useunder{\uline}{\ul}{}
\usepackage{ltablex}

\begin{document}

\title[Activities and Needs of European Fact-checkers as a Basis for Designing HCAI Systems]{Autonomation, Not Automation: Activities and Needs of European Fact-checkers as a Basis for Designing Human-Centered AI Systems}

\author{Andrea Hrckova}
\email{andrea.hrckova@kinit.sk}
\orcid{0000-0001-9312-6451}
\authornotemark[1]
\affiliation{%
  \institution{Kempelen Institute of Intelligent Technologies}
  \city{Bratislava}
  \country{Slovakia}
}

\author{Robert Moro}
\email{robert.moro@kinit.sk}
\orcid{0000-0002-3052-8290}
\affiliation{%
  \institution{Kempelen Institute of Intelligent Technologies}
  \city{Bratislava}
  \country{Slovakia}
}

\author{Ivan Srba}
\email{ivan.srba@kinit.sk}
\orcid{0000-0003-3511-5337}
\affiliation{%
  \institution{Kempelen Institute of Intelligent Technologies}
  \city{Bratislava}
  \country{Slovakia}
}

\author{Jakub Simko}
\email{jakub.simko@kinit.sk}
\orcid{0000-0003-0239-4237}
\affiliation{%
  \institution{Kempelen Institute of Intelligent Technologies}
  \city{Bratislava}
  \country{Slovakia}
}

\author{Maria Bielikova}
\email{maria.bielikova@kinit.sk}
\orcid{0000-0003-4105-3494}
\affiliation{%
  \institution{Kempelen Institute of Intelligent Technologies}
  \city{Bratislava}
  \country{Slovakia}
}

\renewcommand{\shortauthors}{Hrckova, et al.}

\begin{abstract}
To mitigate the negative effects of false information more effectively, the development of Artificial Intelligence~(AI)~systems to assist fact-checkers is needed. Nevertheless, the lack of focus on the needs of these stakeholders results in their limited acceptance and skepticism toward automating the whole fact-checking process. In this study, we conducted semi-structured in-depth interviews with Central European fact-checkers. Their activities and problems were analyzed using iterative content analysis. The most significant problems were validated with a survey of European fact-checkers, in which we collected 24 responses from 20 countries, i.e., 62\% of active European signatories of the International Fact-Checking Network (IFCN). Our contributions include an in-depth examination of the variability of fact-checking work in non-English-speaking regions, which still remained largely uncovered. By aligning them with the knowledge from prior studies, we created conceptual models that help to understand the fact-checking processes. In addition, we mapped our findings on the fact-checkers' activities and needs to the relevant tasks for AI research, while providing a discussion on three AI tasks that were not covered by previous similar studies. The new opportunities identified for AI researchers and developers have implications for the focus of AI research in this domain.
\end{abstract}



\maketitle

\section{Introduction}
A fact-checker is ``a person whose job is to make sure that the facts are correct, especially in something published'' ~\cite{noauthor_fact-checker_nodate}. These professionals work either in larger newspaper agencies (e.g., in AFP, Deutsche Welle, Washington Post) or small or medium-sized NGOs focused just on fact-checking (e.g., Full Fact, PolitiFact.com, FactCheck.org, etc.). Many aspects of the fact-checker's work, as well as the problems this profession faces, differ across the positions (e.g., the required education) or remain unclear (e.g., stages in the fact-checking process; see Fig.~\ref{fig:categ} and geo-cultural differences -- see Fig.~\ref{fig:map}). Nevertheless, research shows~\cite{koliska2024epistemology} that fact-checkers define their role and work as a public service and develop activities to instill a culture of factuality that they value. Exemptions include fringe state‐supported fact‐checkers associated with authoritarian governance structures that weaponize fact‐checking practices and exploit or mimic the social standing of accredited fact‐checkers to fact‐check for national interests~\cite{montana2024fact}.

Research literature reports varying findings on the extent to which fact-checking reduces misinformation, with estimates ranging from minor effects~\cite{clayton_real_2020}, to moderate~\cite{porter_global_2021}, to substantial~\cite{boukes_fighting_2023} reductions in belief in misinformation after exposure to a fact-checking label. In these studies, participants assessed the accuracy of various real and fabricated statements, either with or without a fact-checking tag. Such discrepancies may stem from contextual differences in how misinformation is evaluated, including the type of misinformation, the platform and format of the fact-check, the source of the debunking information, and the metrics used to measure effectiveness (ranging from a 3-point Likert scale on the falsity of the news~\cite{clayton_real_2020} to a 7-point Likert scale on agreement with the news ~\cite{boukes_fighting_2023}), as well as prior beliefs or personality characteristics of the participants. For example, knowledgeable individuals with partisan views tend to scrutinize fact-checking more ~\cite{walter_fact-checking_2020}. Also, who the fact-checker is matters when assessing the success in debunking misperceptions~\cite{van2024fact}. However, in general, fact-checking has been shown to be effective in combination with information literacy interventions in reducing political misinformation in terms of lowering issue agreement and perceived accuracy ~\cite{hameleers_separating_2022}. Fact-checkers also successfully lowered the agreement with attitudinally congruent political misinformation and can help overcome political polarization~\cite{hameleers_misinformation_2020}.

Despite these positive outcomes of fact-checking, active cooperation with social media platforms such as X or Facebook has been ceased recently or is in danger of being ceased despite previous statements that `it works'\footnote{\url{https://web.archive.org/web/20250110130651/https://www.facebook.com/formedia/blog/third-party-fact-checking-how-it-works}}. Without the cooperation with social media, the dissemination of fact-checks will be even less effective. Also, there is currently a trend of replacing the professional fact-checks by community notes, i.e., the consensus of regular users. On one hand, community notes might be an effective approach to mitigate trust issues with simple misinformation flags~\cite{10.1093/pnasnexus/pgae217}. On the other hand, it may result in the fact-checking process being more subjective and susceptible to the bias of community members. More specifically, previous studies showed that community notes tend to be biased towards fact-checking posts from large accounts~\cite{Pilarski_Solovev_Pröllochs_2024}. Also, contextual features -- in particular, the partisanship of the users -- are far more predictive of judgments than the content of the posts and evaluations themselves (e.g., users preferentially challenge content from those with whom they disagree politically)~\cite{10.1145/3491102.3502040}. Finally, a recent study~\cite{borenstein2025communitynotesreplaceprofessional} showed that community notes are more of a complement than a replacement for traditional fact-checking since community moderation relies on professional fact-checking -- more specifically, community notes on posts linked to broader narratives are twice as likely to reference fact-checking sources compared to other sources.

Although growing, the number of professional fact-checkers remains low in comparison to the vast amount of misinformation. The latest census by the Duke Reporters' Lab\footnote{\url{https://reporterslab.org/fact-checking}} identified 378 active fact-checking projects worldwide. Only 88 organizations worldwide (34 in Europe) actively cooperate with social media platforms through the International Fact-checking network\footnote{\url{https://ifcncodeofprinciples.poynter.org/signatories}}, and in some cases, there is merely one fact-checker per country. Imbalance between such scarcity of fact-checkers and misinformation overload causes just a fraction of potential false content is being checked. Furthermore, the work of fact-checkers is laborious and sometimes repetitive. However, there is an opportunity to improve balance by providing appropriate technological support for challenging or repetitive tasks.

While there is an increasing body of research on automation of the fact-checking process, especially with the utilization of Artificial Intelligence (AI) technologies, full automation of the whole fact-checking process is still viewed with skepticism by the fact-checkers as well as researchers~\cite{noauthor_challenges_2020, micallef_true_2022,dierickx_report_2022, juneja_human_2022}. The main reason is that some parts of the fact-checking process require human judgments or actions. Tasks such as assessing the credibility of the sources of evidence are currently not fully automated with sufficient trustworthiness (e.g., by providing appropriate explanations), accuracy, or generability (e.g., checking complex recent claims where evidence may still be missing)~\cite{nakov_automated_2021}. The real potential of AI-based tools, therefore, currently lies in \emph{assisted fact-checking} instead of trying to automate the whole process~\cite{graves_understanding_2016}.

With this in mind, we reinvoke a philosophy of machine design called ``autonomation'', developed in the Toyota Production System~\cite{ohno_toyota_1988}. Autonomation is a blend of ``autonomous'' and ``automation'' and may be described as ``intelligent automation'' or ``automation with a human touch''. Compared to (semi-)automation that is typically orchestrated by a centralized computer controller, autonomation separates the work of humans from machines and thus enables the quick correction of the mistakes made by machines. Yet, autonomation relieves humans of the need to continuously judge whether the operation of the machine is right. In the ``autonomated'' systems, the workers are self-inspecting their work and can source-inspect the work of the machines. This is a difference from an ``automatic'' system (that operates without human intervention) or an ``automated'' system (that requires human input or monitoring but is controlled by technology). Autonomation liberates people from automatable tasks, whereas technology has to serve people and processes, not vice versa. In the context of fact-checking, we use the term metaphorically, as it is more commonly used in the engineering industry.

Although the HCI and information science community has significantly helped to better understand the gap between the social and technical, the existing research works on AI-assisted fact-checking are still often detached from real fact-checkers and thus do not optimally comply with their actual needs and expectations. As Nakov et al. point out, there is a ``\emph{lack of collaboration between researchers and practitioners in defining tasks and developing datasets}'' for automated fact-checking~\cite{nakov_automated_2021}. This contributes to the skepticism of fact-checkers towards automating the entire fact-checking process, especially processes that require human judgments~\cite{noauthor_challenges_2020, micallef_true_2022}. On the contrary, some studies, e.g.,~\cite{10.1145/3242587.3242666} mention too much trust in the system's predictions.

In this study, our objective is to investigate the fact-checking procedures to design the effective and appropriate \emph{human-centered AI tools} to counter false information. The role of humans is central here: the tools should empower humans instead of replacing them. This work offers the following contributions:

\begin{enumerate}
\item
 We unify the various categorizations of the fact-checking process across different computer and social science literature~\cite{noauthor_challenges_2020, nakov_automated_2021, micallef_true_2022, barron-cedeno_checkthat_2020}.

\item
  We investigate the activities, problems, resources, and tools of under-researched European fact-checkers and examine the specifics of fact-checking work in the non-English speaking-regions. We align them with the knowledge from existing studies,  namely~\cite{noauthor_challenges_2020, nakov_automated_2021, dierickx_report_2022, micallef_true_2022}, and visualize them jointly as conceptual models.

\item
 We identify new opportunities for research and development of AI tools designed to support fact-checkers’ daily work in line with human-centered AI principles. In this direction, we discuss three AI tasks that previous studies of fact-checkers’ needs and activities have not addressed yet.
\end{enumerate}

\section{Background and related work}

\subsection{Practices and challenges of fact-checkers in Europe and beyond}
\label{sec:prior_works}

\begin{figure}{
\centering
\includegraphics[width=0.85\linewidth]{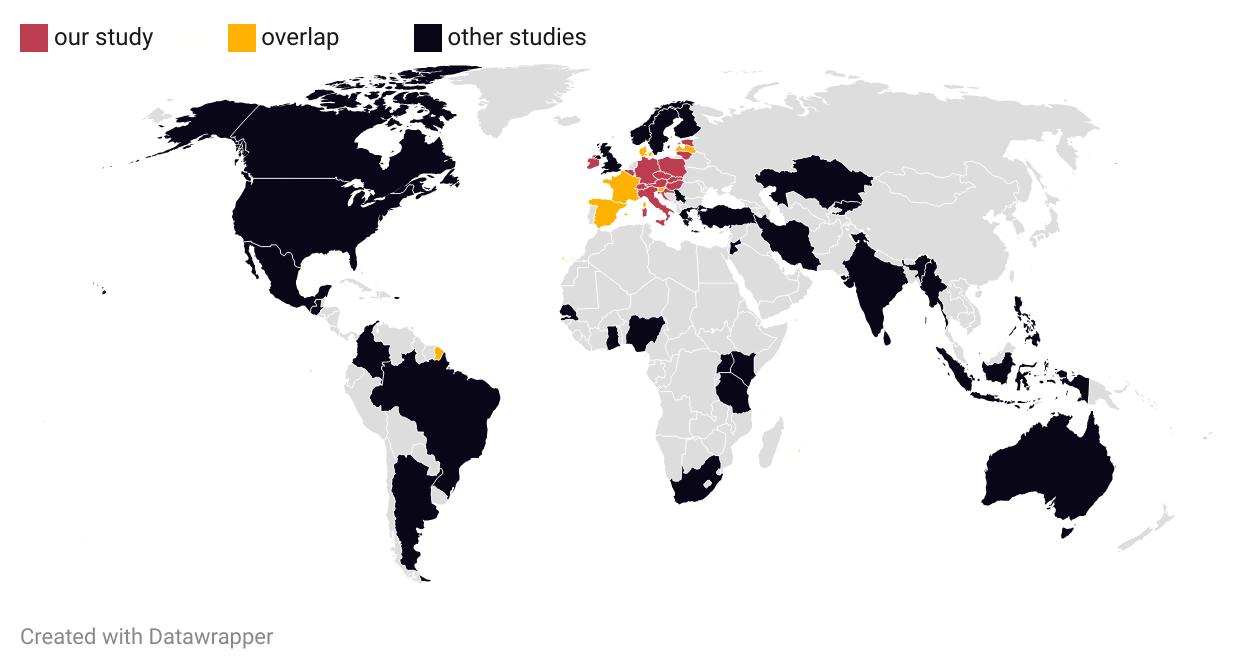}
\caption{The regions covered in existing research on fact-checkers (when specified)}\label{fig:map}
\Description{Four countries overlapping with previous research}
}
\end{figure}

Although the research on fact-checkers' practices and problems is growing (see, e.g., ~\cite{noauthor_challenges_2020, micallef_true_2022, koliska2024epistemology, westlund2024problem}), the European region as well as non-American and non-English speaking fact-checkers remain still under-explored as can be seen on the map (Fig.~\ref{fig:map}). At the same time, there are specific challenges that the European fact-checkers face.

These stem from two main sources. First, the European fact-checkers need to operate in a \textit{highly multinational and multilingual environment}, since many disinformation campaigns easily spread across the borders. To maximize the negative impact in each affected country, disinformation narratives and content itself is often translated and adjusted to specific societal and cultural aspects.

Second, the problem of online disinformation and foreign information manipulations and inference has been \textit{recognized in official policy documents at the level of the European Union (EU) and specific legislation has been drafted} impacting how the social media platforms should respond to disinformation. Specifically, a European approach to tackle these phenomena emphasizes the importance of a transparent, trustworthy, and accountable online ecosystem while promoting quality journalism and media literacy\footnote{\url{https://eur-lex.europa.eu/legal-content/EN/TXT/?uri=celex:52018DC0236}}. Fact-checking has been recognized as one of the important ways to tackle false information and it was supported by various EU initiatives such as the Social Observatory for Disinformation and Social Media Analysis (SOMA)\footnote{\url{https://www.disinfobservatory.org/}} or the European Digital Media Observatory (EDMO)\footnote{\url{https://edmo.eu/}} together with its regional hubs. Most recently, with the adoption of the Digital Service Act (DSA)\footnote{\url{https://commission.europa.eu/strategy-and-policy/priorities-2019-2024/europe-fit-digital-age/digital-services-act_en}}, the previously voluntary Code of Practice on Disinformation\footnote{\url{https://digital-strategy.ec.europa.eu/en/library/2022-strengthened-code-practice-disinformation}} has become the official Code of Conduct, which will be used to determine DSA compliance of the online platforms operating in the EU digital space regarding disinformation risks\footnote{\url{https://digital-strategy.ec.europa.eu/en/library/code-conduct-disinformation}}.

Regarding the existing studies on fact-checkers' practices and problems, the Full Fact report~\cite{noauthor_challenges_2020} was the first one (published in 2020) to globally examine them. Following the semi-structured interviews with fact-checkers from 19 organizations, the report lays out the main challenges fact-checkers face, such as large amounts of potential claims to check, managing tips and suggestions from readers, or the accessibility of information from relevant authorities to check information credibility.

Another study published in 2022 consisted of semi-structured interviews conducted with 21 fact-checkers from 19 countries~\cite{micallef_true_2022}. The results highlight the fact-checkers' motivations, which are driven by a sense of social responsibility, despite encountering numerous challenges throughout the process. This was further reinforced in~\cite{koliska2024epistemology}, which confirms that fact-checkers commonly believe in their ability to determine objective truth, relying on evidence and a transparent process that ensures reproducibility.

Regarding the most common challenges of fact-checkers, these include -- based on~\cite{micallef_true_2022} -- the enormous volume of misinformation, the shortcomings in available tools (such as algorithms that favor virality), confirmation bias of information consumers, the existence of echo chambers in online spaces, and insufficient data and resources for fact-checking efforts. Besides that, the authors point out the reasons for the limited uptake of computational tools by these professionals, such as the limited scope to a specific function within the fact-checking process. Furthermore, the integration of the tools with each other proved to be lacking, as the tools are developed by different entities. The authors suggest developing a unified platform with humans-in-the-loop, or, alternatively, establishing a set of standards to turn the fact-checking process into an efficient and streamlined operation. Five key problematic areas regarding misinformation fact-checking were also identified in~\cite{westlund2024problem} to be: 1) the limited affordances of digital technologies, 2) limited agency on platform infrastructures, 3) limited expertise and human resources, 4) hostility toward fact-checking actors, and 5) fact-checks fueling misinformation.

Additional studies focused on broadening the understanding of various stakeholders (editors, external fact-checkers, in-house fact-checkers, investigators, and researchers, as well as social media managers and advocates)~\cite{juneja_human_2022}  or on the identification of common features of fact-checking methodology~\cite{sanchez2024methodology}.

However, none of these studies focused specifically on the European context. The notable exception is the research on fact-checkers' practices done by NORDIS (NORdic observatory for digital media and information DISorders)~\cite{dierickx_report_2022}, which was, however, limited in scope to Nordic countries, i.e., Norway, Sweden, Finland, and Denmark. Semi-structured interviews with 14 respondents from these countries, supplemented with 5 non-Nordic professional journalists and fact-checkers, were conducted. The findings revealed four types of tools that are needed to automate the parts of the fact-checking process considered ``boring'' for human fact-checkers (social network monitoring, political debate monitoring, claim collection and detection, and context-dependent verification, especially multimedia content) as well as their expected characteristics (e.g., adaptation of tools to Nordic languages). The authors underline the necessity to incorporate the context into the designing process of AI tools, especially 1) the ethical principles of journalism; 2) human values (such as social responsibility to deliver accurate information, creativity, and intuitiveness of the fact-checking process); 3) human expertise with technology; and 4) human relationship to the information resources. In the follow-up study~\cite{dierickx2023automated}, the authors suggest that the focus needs to be moved from a technological point of view toward a social one, provided that a confidence relationship is established between the communities, developers, and fact-checkers/journalists involved from either side of the tool.

In our study, we increase the scope to organizations from 20 European countries and investigate the differences as well as similarities in their problems and needs across various dimensions (including geography, language, or organizational type).

\subsection{AI support of fact-checkers}

A lot of AI research is devoted to assisting fact-checkers by automating
the individual stages of fact-checkers' work~\cite{nakov_automated_2021, zeng_automated_2021}. Reflecting (at least partially) the common steps of the fact-checking process identified in studies above, it focuses on tasks such as finding claims worth fact-checking~\cite{gencheva_context-aware_2017, berendt_factrank_2021, konstantinovskiy_toward_2021}, searching for previously fact-checked claims~\cite{shaar_overview_2021-1, vo_where_2020, sheng_article_2021, kazemi_claim_2021, hardalov_crowdchecked_2022}, or claim verification and evidence retrieval~\cite{zhao_transformer-xh_2020, lee_language_2020, popat_where_2017}. 
Another group of research focuses on the development of the necessary datasets~\cite{shaar_that_2020, nakov_clef-2021_2021, srba_monant_2022, hardalov_crowdchecked_2022} and end-to-end systems or monitoring platforms~\cite{hassan_toward_2017, srba_monant_nodate}. Developing AI-based methods has also been addressed through competitions and data challenges, especially as a part of CheckThat! Labs and SemEval workshops, where tasks such as check-worthiness estimation~\cite{shaar_overview_2021} or detecting previously fact-checked claims~\cite{shaar_overview_2021-1,peng-etal-2025-semeval} have been proposed.

In parallel with research and development of various AI methods and models, several tools aiming to support the fact-checking process have been created. Some of them directly apply AI in order to provide fact-checkers with more advanced features, including automation of some of the fact-checking steps. In the following overview (see also Fig.~\ref{fig:mapping}), we mention especially those that have been explicitly mentioned by the fact-checkers during our study.

At first, there are several tools that allow \textit{social media monitoring}. Typically, these tools are developed for purposes different from fact-checking (e.g., marketing), such as \textit{NewsWhip}\footnote{\url{https://www.newswhip.com/}}, \textit{StoryBoard}\footnote{\url{https://storyboard.news/}} or \textit{BuzzSumo}\footnote{\url{https://buzzsumo.com/}}. Fact-checkers, however, found their way to use these tools also for fact-checking purposes. Some tools are also provided directly by social media platforms, such as \textit{Meta Content Library}\footnote{\url{https://transparency.meta.com/en-gb/researchtools/meta-content-library/}} (a replacement for the former and widely used CrowdTangle tool) dedicated to monitoring Facebook, Instagram, and Threads, or \textit{X Pro}\footnote{\url{https://pro.x.com/}} (a successor of the former TweetDeck tool) dedicated to monitoring X.

For \textit{searching within existing fact-checks}, fact-checkers can also use dedicated tools. To this end, \textit{Google Fact Check Explorer tool}\footnote{\url{https://toolbox.google.com/factcheck/explorer}} allows to search within fact-checking articles that use the so-called \textit{ClaimReview Schema}\footnote{\url{https://schema.org/ClaimReview}}. The ClaimReview Schema allows fact-checkers to annotate their fact-checks by the most important metadata, such as the fact-checked claim, the resulting veracity label, or the link to fact-checked content.

Finally, there are tools that aim to perform \textit{end-to-end fact-checking} (automating all its steps) or at least cover multiple steps at the same time. At first, \textit{Full Fact AI}\footnote{\url{https://fullfact.org/ai/about/}} provides a set of AI-based tools for collecting and monitoring the data, identifying and labeling claims, matching identified claims against a database of past fact-checks, and evidence retrieval to speed up verification of new so-far non-fact-checked claims. Similarly, \textit{ClaimBuster}\footnote{\url{https://idir.uta.edu/claimbuster/}} provides support from monitoring and selecting claims up to their verification against the selected knowledge-bases~\cite{hassan_toward_2017}. Lastly, the \textit{Verification plugin}\footnote{\url{https://www.veraai.eu/posts/verification-plugin}} developed and maintained by the EU-funded vera.ai project (while originally created in the InVID and WeVerify projects), provides a set of tools for detecting various types of false information not only in textual content but also in images and videos. 

Despite existing support and analyses of fact-checkers' work, there is a partial disconnection between the fact-checkers' needs and the efforts to create (AI) tools to support their work. Their design is treated as a technical solution to a technological problem and ignores the complex social and situational context~\cite{juneja_human_2022}. There is also a lack of collaboration between AI researchers and developers with fact-checkers~\cite{nakov_automated_2021}. For example, capabilities of many of the existing fact-checking tools may be limited because various AI-based tools (e.g., detection methods, search engines) typically achieve a worse performance when applied on non-English or low-resource languages. Similarly, the stage \emph{monitoring the online space} is insufficiently covered by AI research in the context of fact-checkers, although it is carried out routinely by these professionals and in some works~\cite{noauthor_challenges_2020, juneja_human_2022} even recognized as ``\emph{the hardest part of the fact-checking process}''. A notable exception is the recent study~\cite{liu2024human}, which indicates a change in trends by using a co-design approach to develop NLP tools for fact-checkers.

This problem is not common only to fact-checkers. Although the conceptual foundations of HCAI are extensively discussed in recent literature~\cite{shneiderman_human-centered_2022, xu_toward_2019} and guidelines for building human-centered AI products exist (e.g.~\cite{google_people_2021}), the industry practices and methods appear to lag behind~\cite{hartikainen_human-centered_2022, bingley_where_2023}. The lack of end-user viewpoint in the early design-related activities is well known from the Human-centered Design (HCD) practice~\cite{hartikainen_human-centered_2022}. Specifically, while tech-savvy fact-checkers are often called for testing of
developed AI tools, these often turn out to be ``irrelevant'' for their
work~\cite{dierickx_report_2022}. This can be prevented by early inclusion of fact-checkers in the tool design process. The challenge is that the capabilities of AI are unclear to users who set the end-user requirements~\cite{hartikainen_human-centered_2022}. This disproportion calls for an interdisciplinary approach, where the practices and problems of users (fact-checkers) are studied by social science and humanities (SSH) researchers and at the same time discussed with AI researchers to design powerful AI tools. 

\section{Methodology}

In this study, we aim to address the identified gap in the field -- our aim is to investigate the activities and needs of fact-checkers to support research and development of the effective and appropriate human-centered AI tools to counter false information. We specifically focus on the European context, which (as shown in Section~\ref{sec:prior_works}) has multiple historical, cultural, language, and legislative specifics; and remains under-studied in this area. Furthermore, the analysis of existing works as well as practical deployment of AI-based tools for fact-checkers showed a need for more human-centered AI approaches. 

Human-centered AI (HCAI) is based on processes that extend user experience design methods such as stakeholder engagement~\cite{shneiderman_human-centered_2022}. The goal is to create tools that augment and enhance human performance. HCAI systems emphasize a high level of human control while embedding high levels of automation that can be achieved by a good design~\cite{shneiderman_human-centered_2022}.
 
Human-Centered Design (HCD) approaches are mentioned to be capable of contributing to the field of HCAI as well~\cite{auernhammer_human-centered_2020}. Specifically, design thinking, previously named Need-Design Response (NDR), focuses on the design and development of any tools or systems for the physical, intellectual, and emotional needs of people. It allows identifying human needs in the early phases of the design project through practices such as need finding. User needs, especially information needs and user interaction with information, focused primarily on the cognitive viewpoint are studied also in information science, specifically in information behavior studies~\cite{bates_information_2009}. 
 
Human-Centered Interaction (HCI), as an interdisciplinary field, adopts a ``human-centered design'' approach to develop computing products that meet user needs, which makes this field a potentially strong contributor to HCAI. Nevertheless, the development of AI systems is still mainly driven by a “technology-centered design” approach  ~\cite{shneiderman_human-centered_2022, xu_toward_2019}. To respond to the challenge, a human-centered AI approach was proposed that places humans at the center of AI design as the ultimate decision makers ~\cite{shneiderman_human-centered_2022, xu_toward_2019}.

The HCAI framework~\cite{xu_toward_2019} includes in addition to ``ethically aligned design'' and ``technology'' also ``human factors design'' to ensure that AI solutions are explainable, comprehensible, useful and usable. To accomplish this, they start from the needs of humans and implement the human-centered design approach advocated by the HCI community (e.g., user research and modeling) in the research and development of AI systems.

However, developers are usually less concerned on what people need in their lives or the social impact of AI~\cite{bingley_where_2023}. These are the defining features of users' positive experiences and need to be addressed for HCAI to be truly human-centered. 

Based on the principles of HCAI and HCD, we make our research study the first stage of the HCAI design process, in which end users are constantly involved in shaping and evaluating the supporting AI tools. This is complementary to the existing human-in-the-loop approaches to fact-checking, where the human workforce is used only to train and validate models in continuous ways, usually to annotate~\cite{demartini2020human} and evaluate models~\cite{yang2021scalable} or provide feedback~\cite{shabani2021sams} in order to reduce bias, increase accuracy, etc.

 \subsection{Research design}
To analyze the activities and needs of fact-checkers, we first engaged these stakeholders individually in \emph{semi-structured in-depth interviews}. Using the iterative \emph{content analysis}, we identified the \emph{activities} -- repeating routines performed by fact-checkers in the individual stages of the fact-checking process. These activities were further validated with previous research work on fact-checker practices and visualized by \emph{conceptual maps} of the fact-checking process.

\begin{figure}{
\centering
\includegraphics[width=0.9\linewidth]{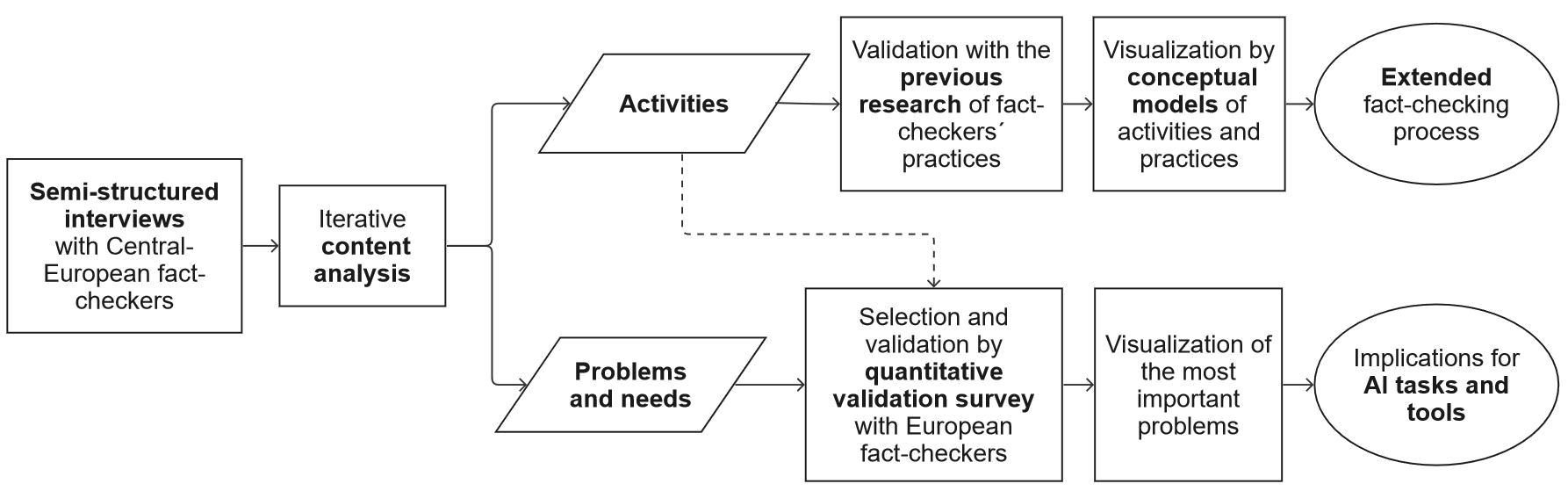}
\caption{Flow chart of the research design (research process, methods, and terminology)}\label{fig:chart}
\Description{The activities were validated with previous research and problems by a survey}
}
\end{figure}

By proceeding from the content analysis, we also identified particular fact-checkers' \emph{needs}. Nevertheless, the needs are to some extent implicit, since they  are inner motivational states to reach goals, and it is possible to be unaware of one' s true needs ~\cite{case_looking_2016}. As not all respondents were tech-savvy, they referred more often to \emph{problems} instead of directly formulating their needs for specific technological support. Therefore, in this paper, we use problems of fact-checkers as a substitution for their needs (i.e., a corresponding need refers to finding a solution that can solve the problem expressed by a fact-checker). The most significant problems (and resulting needs), connected to the activities they relate to, were consequently verified by a \emph{quantitative validation survey}. Finally, the implications for  \emph{AI tasks and tools} were derived. We illustrate this methodology (research process, methods, and terminology) in Fig.~\ref{fig:chart}.

In our research, we address the following research questions:

\begin{itemize}
\item
 \textbf{RQ1:} Which information resources, procedures, and information technologies do fact-checkers use, and how do they depend on local specifics?
\item
 \textbf{RQ2:} What are the biggest problems of European fact-checkers that can be addressed by the support of artificial intelligence?
\item
 \textbf{RQ3:} Which parts of the fact-checking processes are suitable and needed for potential autonomation?
\end{itemize}

To answer RQ1, we conducted semi-structured in-depth interviews, in
which we were interested in how Central European fact-checkers: 1) monitor
the online space to identify potential misinformation; 2) select
potential false claims/narratives; 3) communicate and avoid potential
duplication; 4) verify content credibility and veracity; and finally, 5)
disseminate fact-checks. We compared our findings with fact-checkers worldwide. 

To answer RQ2, we quantitatively surveyed fact-checkers across Europe to assess the weight of the findings of the interviews.

Based on the findings with respect to RQ1 and RQ2 and based on current state-of-the-art research and technical possibilities in the field of AI, we identified the implications for AI development in different stages of the fact-checking process to answer RQ3.

\subsection {Selection criteria for respondents}
The in-depth interviews were conducted with nine fact-checkers from five
major Central European (Slovak, Czech, and Polish) fact-checking
organizations.  All of the respondents performed fact-checking professionally as part of their full-time job; none of them was employed in social media. We involved both external and in-house professionals in management positions (editors-in-chief and project coordinator) and fact-checking-only positions, but also fluid and overlapping fact-checking roles with the roles of journalist, PR manager, editorial manager, and senior research fellow. Our sample represents one of the low-resource language groups, under-explored in AI-research. The sample covers a variety of sizes and types of organizations -- from small NGOs to large news agencies. The majority of organizations also collaborated with social media providers (e.g., Meta). Most of them were part of the IFCN network, which prohibits any kind of political connection. Two organizations were not members of the IFCN (and did not collaborate with social media). These organizations focused on the fact-checking of political discussions in mass media (mainly TV) and partisan news. The product of these two organizations was an article summarizing the main arguments in a misleading topic of interest (not a structured fact-check of a claim). 

For this work, we define Central Europe as the countries of the Visegrád Group~\cite{tiersky_europe_2004}. While Central Europe geographically belongs to the European region, its historical and linguistic context differs from Western Europe (Eastern Block legacy, Slavic languages). This has many consequences on its current social and cultural setting and related issues (e.g., the level of Russian influence). Contextual factors affect the choices of action and the use of sources and channels in the online environment~\cite{agarwal_exploring_2018}. We can thus assume that some needs of Central European fact-checkers can differ from the fact-checkers elsewhere in Europe/the world. In particular, we can expect differences when compared with high-resource language areas.

Nevertheless, Central European fact-checkers form a sample too small to derive conclusions about the problems of fact-checkers (not covered in existing literature). Therefore, we validated the results of the interviews with a validation survey.
The survey was responded to by 24 representatives (N = 24) of 21 European fact-checking organizations, covering 20 countries. Albeit a nominally small sample, our survey respondents represent approximately 62\% of active European fact-checking organizations that are IFCN signatories.

All fact-checkers participating in the survey had to be members of recognized professional groups.
Therefore, organizations from CEDMO (Central European Digital Media Observatory) hub\footnote{\url{https://cedmohub.eu/}} were selected for the in-depth interviews. The survey was directed at EDMO (European Digital Media Observatory) hubs\footnote{\url{https://edmo.eu/edmo-hubs/}} members and to the International Fact-Checking Network (IFCN's)\footnote{\url{https://ifcncodeofprinciples.poynter.org/signatories}} signatories operating in Europe. The survey respondents were recruited through contacts provided by the EDMO or LinkedIn Premium account. Both external fact-checkers and in-house professionals were involved, but none of them was employed in social media. The position names of the respondents mentioned in their LinkedIn profiles were fact-checker, verification officer, journalist, redactor, or analyst in the fact-checking parts of the organizations.

\subsection{Method 1: Interviews with fact-checkers}

The semi-structured interviews were conducted with one organization at a time, and each required between one and one and a half hours. The interviews were conducted from December 2021 to February 2022 using a video conference system using the English, Slovak, or Czech language. 

\begin{figure}{
\centering
\includegraphics[width=\linewidth]{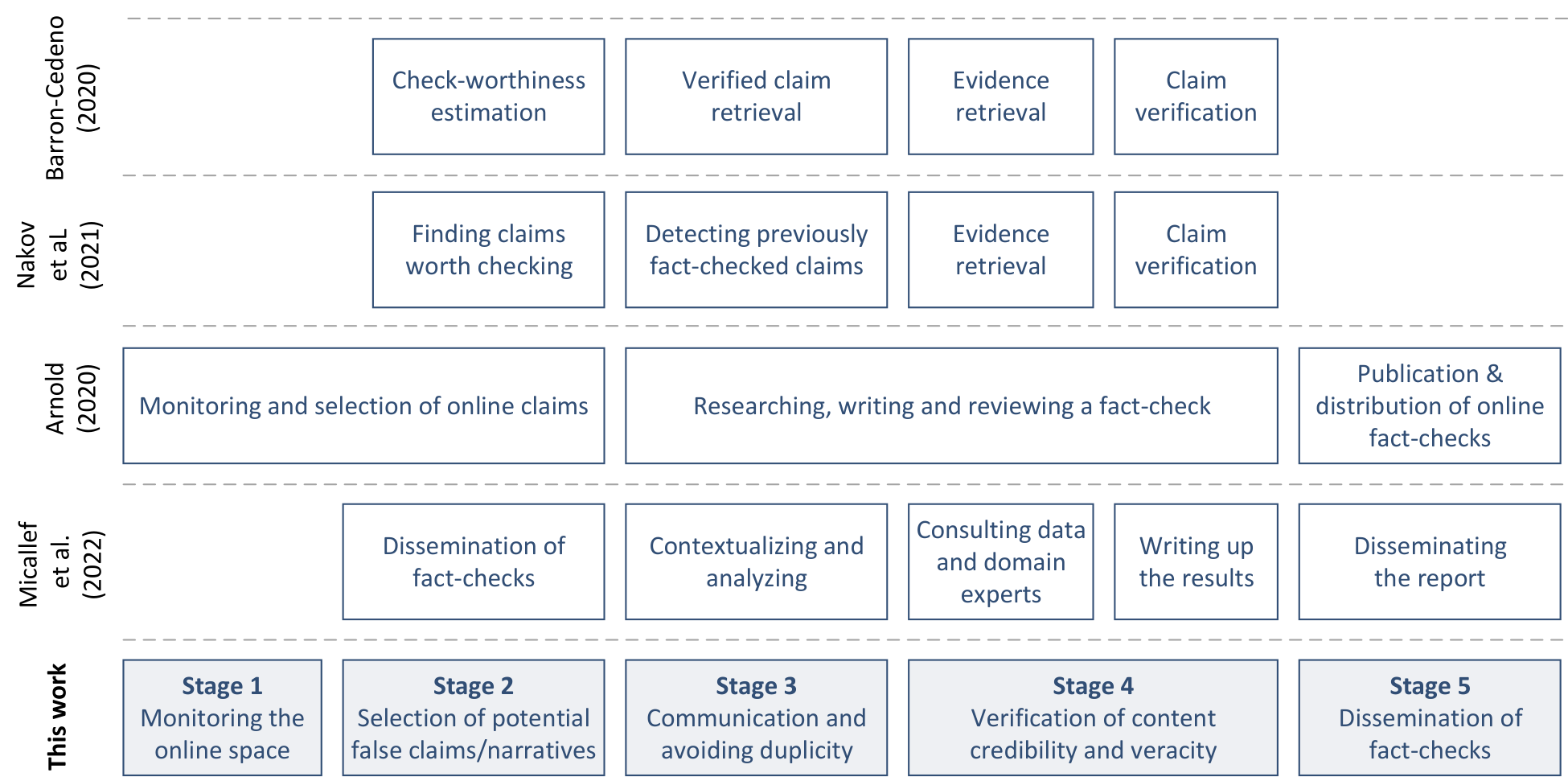}
\caption{Various categorizations of stages in the fact-checking process in the available research}\label{fig:categ}
\Description{We unified the categorizations of fact-checking process to five stages}
}
\end{figure}

When designing the interview, we faced many ambiguities in the fact-checking process. The number of stages, as well as terms used to name them, varies across the literature and tends to depend on the view that the research is dedicated to (social
sciences versus AI) as shown in Fig.~\ref{fig:categ}. Existing work in AI typically recognized four core fact-checking stages, while the first study of today's practices of fact-checkers~\cite{noauthor_challenges_2020} mentioned three different stages. Therefore, we were obliged to redefine the stages found in the literature when designing the interview questions.

When designing the interview questions for respondents, we focused on the human-information interaction processes. Asking about information resources to identify user needs is a very common approach in information science research ~\cite{case_looking_2016}. We also included the interview questions about problematic, time-consuming, and repetitive work, which are common in software requirements specification. The full listing of interview questions can be found in Annex A. 

The interviews transcripts were read and analyzed multiple times using iterative content analysis. The techniques we have employed include open, axial, and selective coding. All codes are represented in conceptual models (Fig. \ref{fig:monitoring}-\ref{fig:dissemination}).

\subsection{ Method 2: Validation survey}
 The interviews yielded many problems as well as some implicit inputs on what is desired and needed for fact-checkers to be more efficient in tackling disinformation. These inputs were discussed, using brainstorming methods, and problems that were most cognitively demanding, tedious, time-demanding, and complex, as well as brought inconveniences and uncertainty with them, were selected. Finally, 13 problems were chosen for the validation survey based on the user activities and current technical feasibility, resource feasibility, and sustainability of potential AI solutions (tools) (see the validation survey questions in Annex B).

The survey also served to validate the individual responses; therefore, for each activity, the respondents were queried on:

\begin{itemize}
\item
 how often fact-checkers need to be involved in such activity (on a  six point Likert scale ranging from ``more times a day''=6 to
 ``never''=1) and~ 
\item
 how difficult the activity is (on a five point Likert scale ranging from  ``No problem, I like to do it''=1 to ``I perceive it as a big problem. I really need help with this''=5).
\end{itemize}

 To identify the most pressing problems that need automation, we calculate their importance as follows.

$$C(a_i)= med(f_i) \times med(l_i)$$

 where $C(a_i)$ is an importance coefficient that quantifies the need for automation within activity $i$, $med(f_i)$ is a median score of the frequency of the fact-checkers' involvement in activity~$i$ and $med(l_i)$ is a median score of the level of difficulty of activity~$i$. As such, the most severe problem would get a coefficient of 30, and the least important problem a coefficient of 1.

The quantitative validation survey was conducted from March 2022 to May 2022.

\subsection{Results presentation}

The results of the interviews, i.e., the details of the fact-checking process, based on the fact-checkers' answers in the interviews, are summarized and visualized with conceptual models using CmapTools\footnote{\url{https://cmap.ihmc.us/}}. They are also compared with the fact-checking process of the organizations in~\cite{micallef_true_2022}, fact-checkers' needs in~\cite{noauthor_challenges_2020, juneja_human_2022}, and utilized tools in~\cite{dierickx_report_2022}. Differences are outlined in different fonts according to the legend under each figure.

The identified problems may vary across the different tasks that fact-checkers are involved in. This may be considered a limitation of our study and prompts future research focused on complex process analyses of individual fact-checking organizations.

\section{Results and findings: Interviews with fact-checkers}

In general, regarding RQ1, the results of the interviews indicate that the fact-checking process in Central Europe is similar to that in the rest of the world. However, since the fact-checking process differs in the various research studies (see Fig. \ref{fig:categ}) and some of the stages were missing according the results of our content analysis, we \textbf{renamed and extended its stages from the previous works} (see Fig. \ref{fig:categ}). Stage 2 (Selection of potential false claims/narratives) and Stage 4 (Verification of content credibility and veracity, in computer science labeled as evidence retrieval and
claim verification) were inspired by~\cite{barron-cedeno_checkthat_2020, nakov_automated_2021}, but based on the interviews with fact-checkers, the names of the stages were renamed (extended) following these findings: 1) fact-checkers do not select only claims, but also narratives, 2) fact-checkers verify the sources of claims besides individual claims (collectively referred to as content). 

We expanded the fact-checking process as it has been understood in previous AI research in: Stage 1 (Monitoring the online space), Stage 3 (Communication and avoiding duplication), and Stage 5 (Dissemination of fact-checks). We also extended the dissemination part of the process: no longer it means only the publication and marketing of fact-checks, but includes a novel subcategory -- the identification of the other versions (i.e., other appearances in the online content) of the same misinformation. We use the term ``misinformation'', as not every respondent was fact-checking at the level of claims: some of them were interested in repetitive topics (narratives), whereas their final product was a fact-checking article (not a structured fact-check of a claim). This is in accordance with two identified types of fact-checking~\cite{juneja_human_2022}: 1) short-term claims centric and 2) long-term advocacy centric.

As we found out, some identified problems were specific to the investigated region and connected mainly to the capacity and financial reasons (e.g., weaker collaboration with other fact-checkers or dissemination in an attractive form) and limited language capabilities of the tools (e.g., for monitoring potential misinformation or for finding the previously published fact-checks). The virtue of Central European fact-checkers was ingenuity when forced to use the freely available tools. Its manifestation could be seen in the use of personalization on social networks using fake profiles to monitor misinformation, but also in the creative use of CrowdTangle to monitor notorious liars. Below we present a detailed description of the process and problems of our respondents in comparison with the fact-checkers around the world. 

\subsection{Monitoring the online space}

\begin{figure}{
\centering
\includegraphics[width=\linewidth]{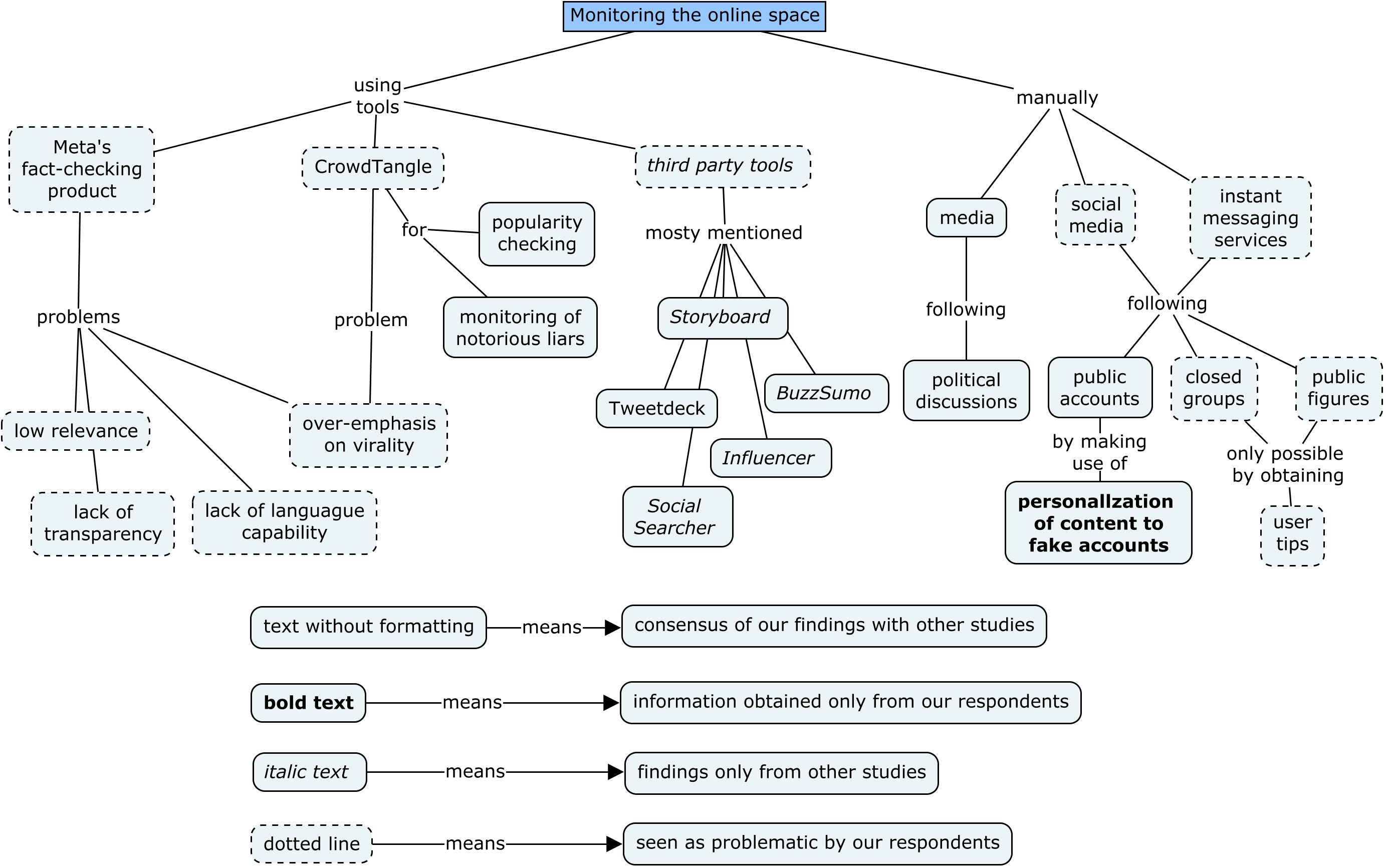}
\caption{Conceptual model of the monitoring part of the fact-checking process}\label{fig:monitoring}
\Description{The tools, activities and problems of fact-checkers with monitoring potential misinformation}
}
\end{figure}

Monitoring (Fig.~\ref{fig:monitoring}) is globally seen as one of the hardest parts of the fact-checking process. The process was illustrated, for example, by P1:

\begin{quote}
\emph{``Monitoring of online space is done in two ways: manually or automated. When authors browse media manually (the classic media such as the New York Times and the Washington Post), they search for topics that are interesting for our audience.  Some of us also do automated monitoring in CrowdTangle... I do that every morning by coffee, where I check notorious sources -- some lists have about 800 [social media] pages.``}
\end{quote}

Monitoring is a part of the process that fact-checkers are the most open to autonomating~\cite{dierickx_report_2022}. Still, there is just a limited pool of tools that are a bit helpful for Central European as well as other non-English-speaking fact-checkers on social media.

For Meta's platforms (Facebook, Instagram), fact-checkers originally used to rely on CrowdTangle. As this tool was originally developed for marketing purposes, it has put overemphasis on virality, which was not always useful for fact-checkers. This was mentioned by our respondents as well as other fact-checkers around the world~\cite{noauthor_challenges_2020, dierickx_report_2022, micallef_true_2022}. Despite this, fact-checkers agreed on the usefulness of this tool for creating and monitoring lists of notorious liars. It was also useful for checking the popularity of misinformation, for example, in cases when it was reported by regular users. Despite a clear value provided to fact-checkers, Meta abandoned its active development at first, and in August 2024 it was deprecated completely. From that time, fact-checkers are provided with a replacement tool built on the top of new Meta Content Library (MCL), which, however, does not provide the same set of features as the original CrowdTangle tool and which are recognized as missing ones by fact-checkers.

In the case of the original Twitter platform, fact-checkers relied on a simple but efficient TweetDeck tool. Its current version, branded as X Pro, became in July 2023 a paid proprietary tool that is available only to verified and premium users of the X platform, which reduced its availability and usefulness for fact-checkers.

In Norway, fact-checkers rather rely on their Storyboard tool to find problematic content~\cite{dierickx_report_2022}, in other research ~\cite{juneja_human_2022} BuzzSumo, Social searcher or Influencer are mentioned.

Fact-checkers involved in Meta's \emph{Third-Party Fact-Checking Program} have access to Meta's fact-checking tool that gathers reported data from users and their own detected potential misinformation. Although it is the only fact-checking product from social media, the fact-checkers from Central Europe mention its lower relevance for the Slavic language group. This is also similar to the experience of some other fact-checkers from non-English speaking countries~\cite{noauthor_challenges_2020}. Some of the Central European fact-checkers bypass the problem with Meta's fact-checking tool by creating fake profiles that follow suspicious groups, pages, and users. This way, they can take advantage of the sophisticated personalization of the platform to recommend other questionable sources more efficiently in contrast to the official Meta's tool for fact-checkers.

The resources that fact-checkers monitor depend on the aims of organizations (mass media vs. social media). It usually reflects the current importance of a particular social medium in a particular country~\cite{micallef_true_2022, noauthor_challenges_2020}. However, in Central Europe, it involved only Facebook and WhatsApp for capacity reasons. The ideal monitoring requires substantial manual effort, as fact-checkers need to access information across a multitude of public and private social media channels (e.g., a lot of misinformation comes screenshotted from Telegram to Facebook). Although instant messaging services are heavily used, currently there are no autonomated means to monitor them because of legal constraints (e.g., troll channels are still private channels), technological constraints (e.g., limited or no API is provided for this purpose), and capacity constraints (fact-checkers cannot resolve all of the user tips). User tips are nowadays the only relevant source for monitoring instant messaging services. Therefore, Central European fact-checkers still focus on open social media accounts.

\subsection{Selection of potential false claims/narratives}

\begin{figure}{
\centering
\includegraphics[width=0.9\linewidth]{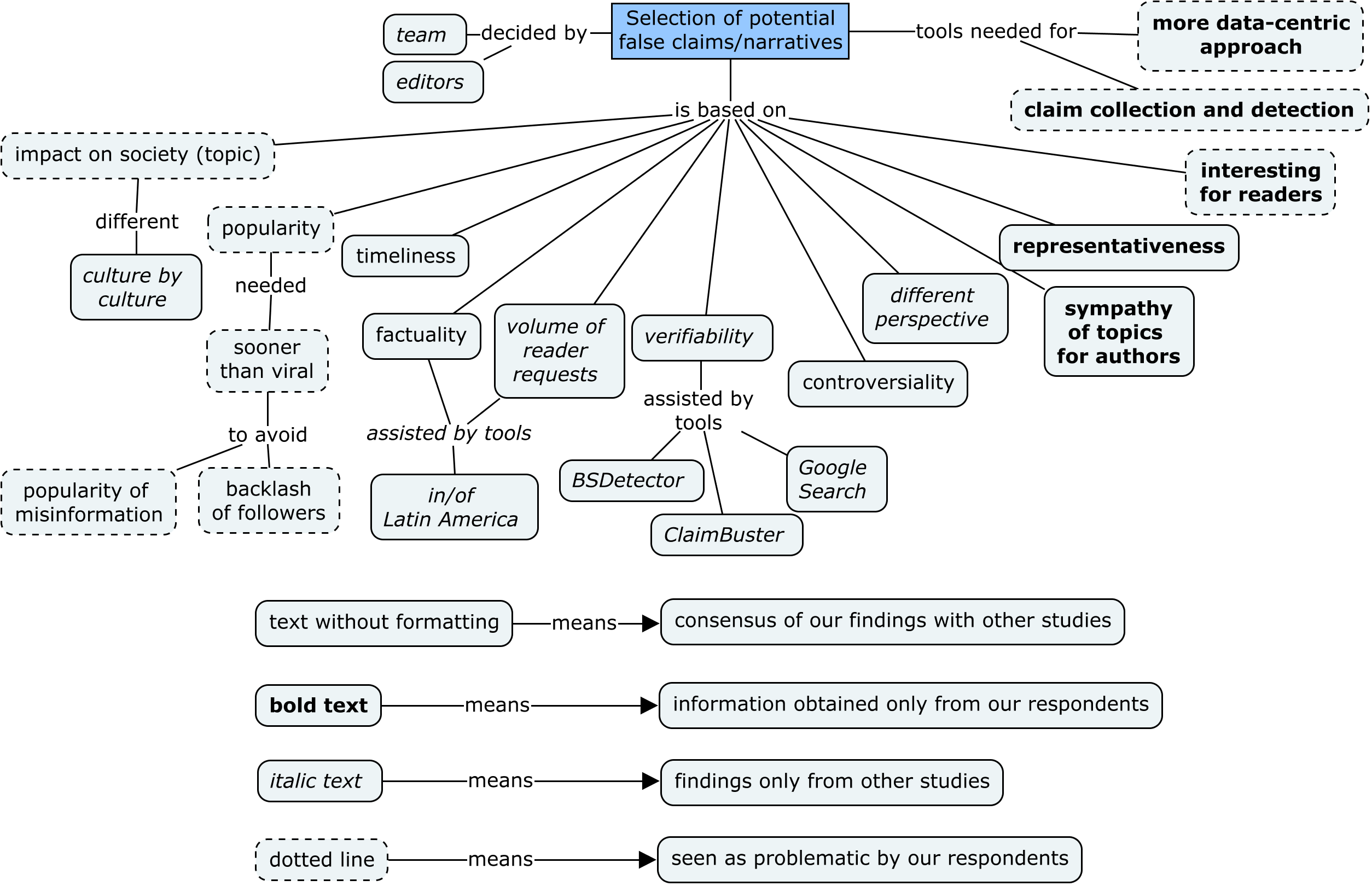}
\caption{Conceptual model of the selection of potential misinformation}\label{fig:selection}
\Description{The tools, activities and problems of fact-checkers with selecting the misinformation to fact-check}
}
\end{figure}

Our participants spend considerable time filtering and prioritizing content for their audience (Fig.~\ref{fig:selection}). This is in line with previous research~\cite{noauthor_challenges_2020, dierickx_report_2022, micallef_true_2022, juneja_human_2022}. The selection is done either by an editorial team in larger organizations or collectively as a team in small organizations (also in~\cite{micallef_true_2022}). In addition to that, we noticed some subjectivity in the selection of claims. None of the responding organizations had general guidelines for the misinformation selection process. However, fact-checkers agreed on some informal selection criteria for misinformation, considering the impact on their audience and society. The popularity or virality of potential misinformation was mentioned only as one of the factors for selection. Most of the respondents would appreciate early virality warnings on misinformation, as put out by one of the fact-checkers (P4):

\begin{quote}
\emph{``We need a tool that would alert us when something starts to spread before it becomes viral\ldots{} At the moment, we look more at how posts on social media impact society or discussion.''}
\end{quote}

Despite its importance, the impact on society was estimated only intuitively in the interviewed fact-checking organizations. Remarkably, most of them did not analyze data about the impact of their fact-checks on their audience because of the capacity constraints. Some insights based on social media engagement would therefore be helpful for a more data-based approach for filtering and selection of potential false claims/narratives. Furthermore, there are many cultural differences in the selection of claims topics~\cite{micallef_true_2022}. Therefore, a (meta)research of the locally important narratives, enriching the data-based approach is needed. 

Some Central European fact-checkers also mentioned timeliness, factuality, representativeness, controversiality of potential
misinformation, or sympathy of the topic to the authors as factors for selection. All these factors are consistent with previous research~\cite{micallef_true_2022}, as well as putting more emphasis on societal impact than virality, including issues that affect minorities or disadvantaged groups. In comparison to our respondents, the volume of reader requests, verifiability, and the importance of providing a different perspective for their readers were mentioned as important factors for prioritization in previous research~\cite{micallef_true_2022}. Some respondents in this research preferred to fact-check influencers, while others were aware of the high probability of a backlash from their followers and preferred not to check. Note that fact-checkers all over the world face legal actions. Some even reported facing personal danger, even physical attacks~\cite{noauthor_challenges_2020, juneja_human_2022}. The latter has not been the case of our respondents, but it underlines a need for a tool that alerts fact-checkers before something goes viral.

Checking whether a claim is verifiable is one of the tasks that the fact-checkers are open to automating the most~\cite{dierickx_report_2022}. In some countries, it is performed by a simple Google search or by using freely available tools ~\cite{dierickx_report_2022} like ClaimBuster~\cite{hassan_toward_2017}, BSDetector~\cite{micallef_true_2022} or Full Fact AI. 
There are also open-source tools developed and used by fact-checking organizations in Latin America that: 1) differentiate
between the opinion and factual statements~\cite{beers_examining_2019}; and 2) optimize the process of collecting and responding to WhatsApp reader requests~\cite{bhuiyan_investigating_2020}. None of these tools was mentioned or
utilized by our interviewees. Study \cite{dierickx_report_2022} mentions five reasons for not adopting AI tools by fact-checkers: 1) lack of knowledge about the existence of such tools; 2) lack of time to learn the tool; 3) lack of resources to pay for the tool; 4) lack of knowledge of how the AI tool works; and 5) difficulty in integrating the tool into the organization's workflow. Keeping fact-checkers in the design process of the AI tools and appropriate training may help overcome these barriers~\cite{dierickx_report_2022}.

\subsection{Communication and avoiding duplication}

\begin{figure}{
\centering
\includegraphics[width=0.75\linewidth]{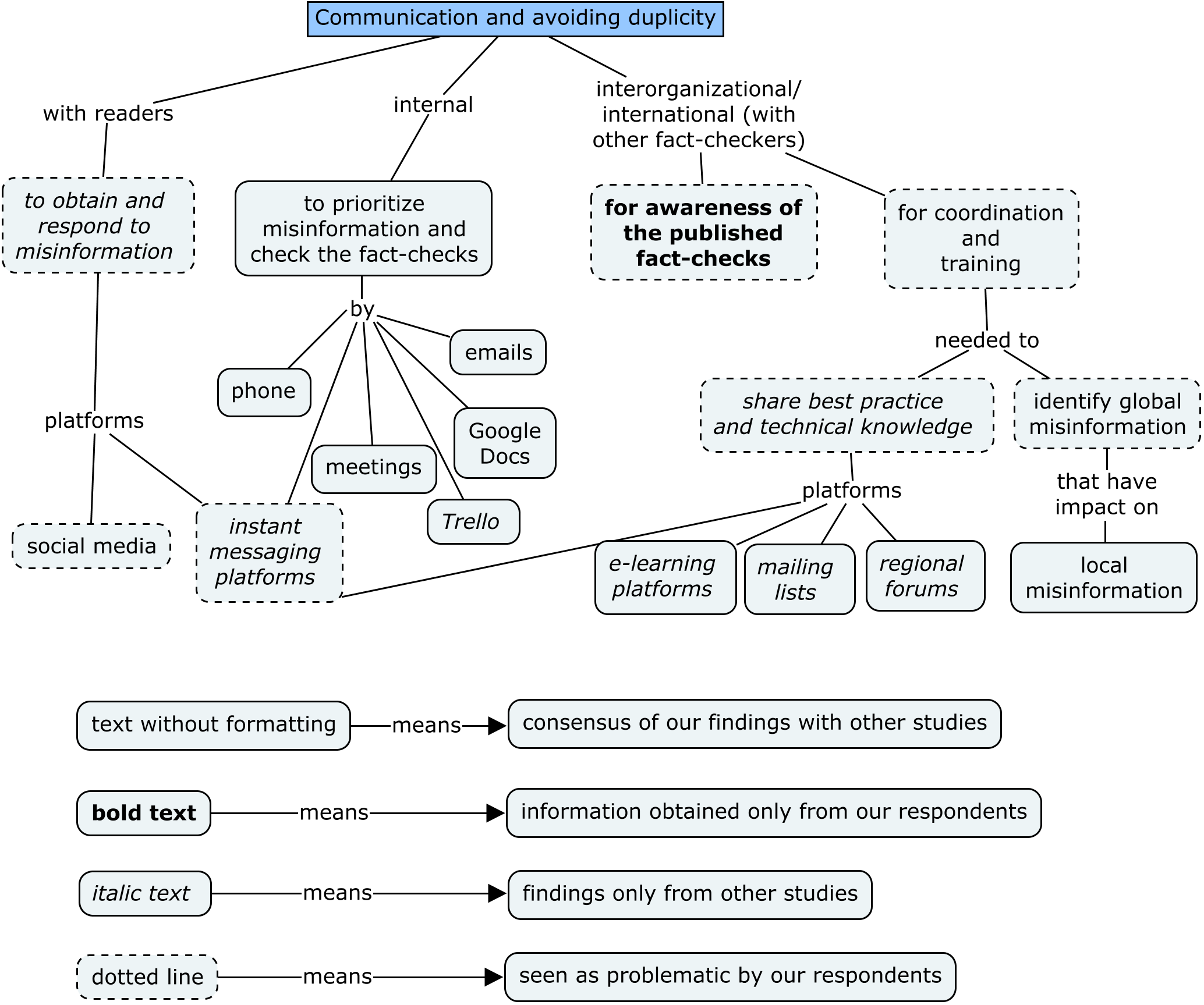}
\caption{Conceptual model of the communication part of the fact-checking process}\label{fig:communication}
\Description{The tools, activities and problems of fact-checkers in the communication part of the process}
}
\end{figure}

Communication within fact-checking teams (Fig.~\ref{fig:communication}) is generally not perceived as problematic. Teams typically rely on traditional communication channels and freely available tools such as Google Docs and instant messaging services. However, interorganizational—and especially international—communication is considered crucial, as fact-checkers need: 1) to stay informed about fact-checks currently being verified or published, and 2) to coordinate efforts with colleagues worldwide.

It is precisely this global coordination (Fig.~\ref{fig:communication}) that respondents identified as a challenge. While fact-checkers actively search for fact-checks from other organizations, they are usually aware of only a fraction of those published. Improving communication or implementing autonomated systems for identifying misinformation across fact-checking and journalistic organizations could accelerate also the identification of local misinformation, as one fact-checker (P2) explained:

\begin{quote}
\emph{``As an international team, we {[}can internally{]} share information {[}about misinformation in individual countries{]}. {[}In this way{]}, we can predict that the same misinformation will appear {[}in our country as well{]}. It can originate on the opposite side of the world, or many times the fake news comes from Germany; we know that there is a probability that it will appear to us, too, and mostly we are right.''}
\end{quote}

Collaboration among fact-checking organizations is widely recognized as essential for sharing tips and technical expertise~\cite{micallef_true_2022}. In some regions, such as the Balkans or Latin America, it is common practice to exchange information through mailing lists, regional forums, or WhatsApp groups dedicated to fact-checking~\cite{micallef_true_2022, juneja_human_2022}. Some initiatives, such as Factify, have even developed e-learning platforms with video tutorials exclusively for fact-checkers~\cite{juneja_human_2022}. By contrast, Central European fact-checkers—often working in small NGOs—rarely engage in such exchanges. Strengthening collaboration, building coalitions, and developing shared training opportunities are key areas for improvement in the region to enhance the capacity and influence of fact-checkers. 

\subsection{Verification of content credibility and veracity}

Content evaluation is widely regarded as an activity that requires human judgment. Both in our region and globally~\cite{noauthor_challenges_2020, micallef_true_2022, juneja_human_2022}, this process depends heavily on communication with people. Fact-checkers view domain experts as valuable sources of information, and at times, they contact the authors of disputed content to request evidence or responses to fact-checks. They also engage with government officials and press agencies to request or clarify data—a step that can significantly delay the process, as such officials tend to be highly cautious in their statements. This reliance on interpersonal communication and expert input underscores the complexity of content evaluation and represents a key obstacle to its automation.

\begin{figure}{
\centering
\includegraphics[width=\linewidth]{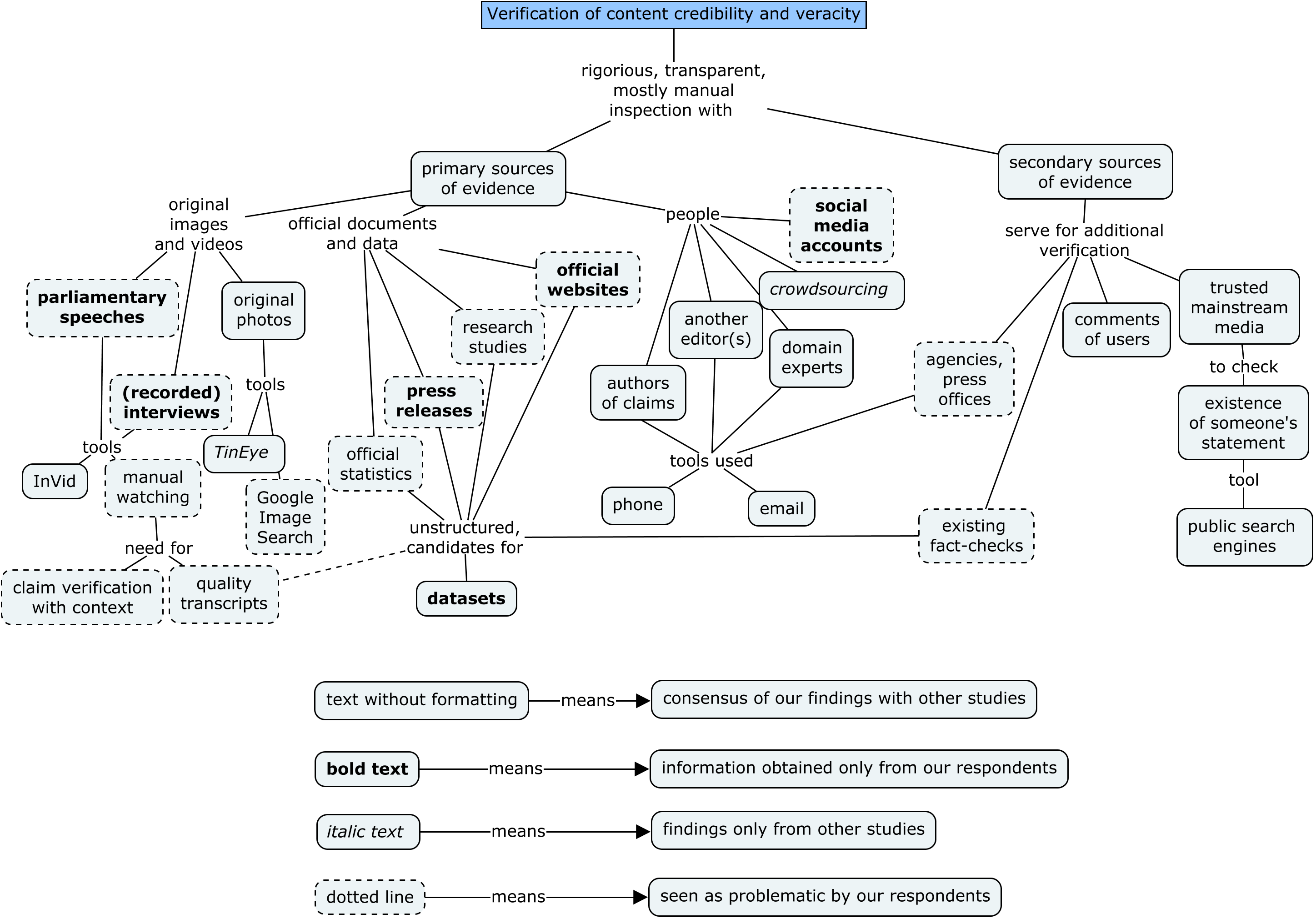}
\caption{Conceptual model of the verification part of the fact-checking
process}\label{fig:credibility-veracity}
\Description{The tools, activities and problems of fact-checkers in the verification part of the
process}
}
\end{figure}

Nevertheless, while evaluation itself resists full automation, fact-checkers view \emph{evidence retrieval}—a time- and labor-intensive stage—as a promising candidate for technological support and partial automation. Evidence retrieval involves the acquisition of \emph{primary
sources} to achieve the maximum possible
objectivity (Fig.~\ref{fig:credibility-veracity}). Our focus on human-information interaction enabled us to go deeper into the specific resources that the fact-checkers use for evidence retrieval. The textual sources involve mostly official data
(as statistics), official information on the websites of governments,
private companies, and NGOs, or original research studies. A
fact-checker (P3) describes the process:

\begin{quote}
\emph{``We have to find the right source for the data, such as in the Polish statistical bureau, central controller's office, Eurostat\ldots{} If the claim is not about data, we try to find facts or proof since as journalists, we can be sued for everything we write\ldots{} It is not just debunking whether {[}the claim{]} is true or not\ldots{} Lots of claims are manipulative or partially true, \ldots{} Very often we use experts (such as in finance, energy, or climate) to evaluate~ information in statements -- including implicit ones... The media is not our sources, they are good for background or if someone said something in the media.''}
\end{quote}

The official data are hard to retrieve as they are not usually in easily accessible formats or are not publicly available.
Fact-checkers in all regions experience different difficulties in reaching primary and official data sources, depending on the level of development of a country~\cite{micallef_true_2022}. However, not being able to verify a claim is one of the biggest frustrations of fact-checkers~\cite{dierickx_report_2022}.

Fact-checkers often work with audiovisual evidence, such as
parliamentary speeches or recorded interviews to check whether a person mentioned the claim that someone else accuses him or her of saying. The fact-checkers frequently require watching hours of videos, many times without any result. Therefore, most of the Central European respondents would appreciate having searchable transcripts of the videos in local languages.

Secondary sources such as mass media are rarely used for verification of claims. They were mentioned (if at all) rather as
illustrative than reliable. Sources like Wikipedia were never used or considered credible for verification during fact-checking, as confirmed also by~\cite{dierickx_report_2022}. Nevertheless, according to~\cite{guo_survey_2022}, textual sources, such as news articles, academic papers, and Wikipedia documents, have been one of the most commonly used types of evidence for automated fact-checking. As the above mentioned sources are not utilized by fact-checkers, verification tools trained or tested just on Wikipedia and mass media datasets will never be sufficient for the work of these professionals.

The largest exception to the non-usage of secondary sources are existing fact-checks of other fact-checking organizations. These are seen as the most credible secondary source of evidence for Central European fact-checkers. Existing fact-checks within the same organization are considered good sources of information, as politicians and other actors often repeat themselves. However, previously existing fact-checks (regardless of their relative provenience) are often hard to retrieve as P3 continues:

\begin{quote}
\emph{``The most cumbersome is to find, if a claim or statement has already been fact-checked or if fake news was verified by someone else. Most of the fact-checks come from abroad. We have to check if English, French, Italian, German fact-checkers did it… If yes, we sometimes use it so that we can prove the history of spreading, then we quote these fact-checking websites.``} 
\end{quote}

Another reason is, that many fact-checks are not created with the necessary structure or metadata and are thus difficult to search through services such as Google Fact-Check Explorer. Adhering to the most common fact-check schema, the ClaimReview, is rather cumbersome for fact-checkers under their present technological conditions.

According to our respondents (Fig.~\ref{fig:credibility-veracity}), few search engines are helpful for fact-checkers. They mention mostly the Google Reverse Image Search that can identify similar images. Ordinary Google Search is not sufficient for the needs of Central European fact-checkers as it does not retrieve the most credible and hard-to-search textual sources of evidence. This is in accordance with~\cite{micallef_true_2022} as well as~\cite{hasanain_studying_2022}, who studied the effectiveness of commercial web search engines for fact-checking. They found out that the engine's performance in retrieving relevant evidence was weakly correlated with the retrieval of topically relevant pages.

InVid tool in the Verification plugin was seen as very relevant for our respondents as it decomposes the videos into keyframes (images) that are possible to be reverse-searched. Nevertheless, as some misinformation is very complex, fact-checkers need to do more than reverse-search the images. Interestingly, participants in the study of~\cite{micallef_true_2022}, who reside mostly outside of Europe, do not report being aware of these tools.

Content verification is seen by our respondents as an ``intuitive process'' that results from the long-term practice of fact-checkers,
often journalists. Fact-checkers prevailingly stated that they notice questionable sources ``at first sight''. However, one respondent shared his credibility indicators for the evaluation of sources, namely: non-existent author, problematic source (in the past), factuality, distortions, biases, objectivity, out-of-the-context use, sentiment and argumentation fallacies.

In previous research \cite{micallef_true_2022} a ``contextualization stage'' was mentioned,
where most fact-checkers looked into the evolution of the claim from its origin to the present state to help readers understand their conclusions. This process is very challenging, especially in foreign languages. One of our interviewees mentioned the problematic tracing of such a history, particularly on social media, where the search capabilities are very limited.

Fact-checkers’ verification processes are generally very transparent, as the websites of fact-checking organizations often explain how the judgment was reached, attaching the sources as well (often their archived versions). Besides that, it is very rigorous. As reported, none of the fact-checks is published without consulting it with at least one editor. Participants from Africa and Oceania have even longer peer review processes than Western countries~\cite{micallef_true_2022}.

\subsection{Dissemination of fact-checks}

\begin{figure}{
\centering
\includegraphics[width=0.85\linewidth]{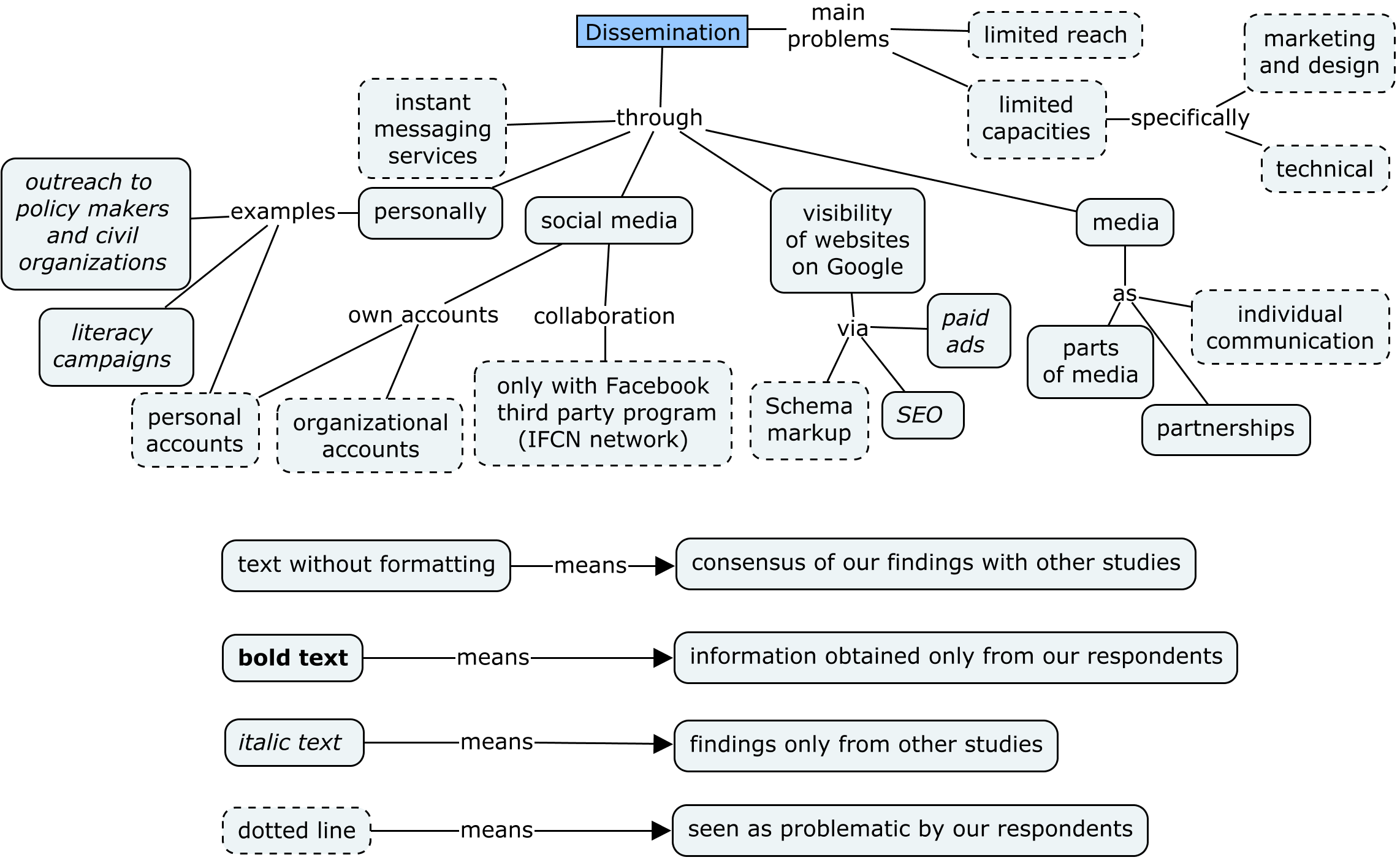}
\caption{Conceptual model of the dissemination part of the fact-checking process}\label{fig:dissemination}
\Description{The tools, activities and problems of fact-checkers in the dissemination part of the process}
}
\end{figure}

Because of capacity reasons, dissemination of fact-checks mostly takes place where the misinformation spreads -- on social media, mostly Facebook (Fig.~\ref{fig:dissemination}). Fact-checks are published on the websites of the organizations. In connection to web publishing, Central European fact-checkers frequently report limited technological capabilities in their organizations (unlike some fact-checkers throughout the world~\cite{noauthor_challenges_2020, micallef_true_2022}). These prevent them from full exploitation of search engine optimization (SEO), ClaimReview, appearing in Google News, or paid ads on Google as noted by P5:

\begin{quote}
\emph{``We thought of using the ClaimReview format, we tried it on fifteen claims in the Google Search Console. But as there are many fact-checks, we would need a process that would fill it automatically. It was very time consuming to fill all those fields.''}
\end{quote}

Our respondents from small NGOs also struggle with personal (marketing and design) capacities to be able to make their fact-checks more appealing and popular by including visuals (e.g., infographics, videos, comic stripes). This is a common practice for copy editors in some fact-checking organizations ~\cite{juneja_human_2022}. Nevertheless, fact-checkers communicate with the media, as P6 claimed:

\begin{quote}
\emph{``Sometimes the media notice that we fact-checked something, sometimes we reach the media ourselves, sometimes we have a project with them to fact-check something.''}
\end{quote}

Fact-checkers around the world pointed out that they communicate with their audience through instant messaging services, particularly WhatsApp~\cite{noauthor_challenges_2020, micallef_true_2022, juneja_human_2022}. Nevertheless, this communication is more cumbersome, and Central European fact-checkers do not utilize these channels.

Some of our fact-checkers, as well as the respondents in the study of~\cite{micallef_true_2022} and~\cite{dierickx_report_2022} revealed concerns about the limited reach and potential of their outcomes. The collaboration with social media platforms achieved some success. Nonetheless, fact-checking is a long process, and carrying out all its stages allows misinformation to spread in the meantime. Unlike our respondents, some fact-checking institutions also reach out to policy makers and civil organizations or organize literacy campaigns to strengthen the dissemination of facts.

\section{Results and findings: Validation survey}

 \subsection{The problems of European fact-checkers}
 In general, we can conclude that the validation survey supports and extends the findings of the in-depth interviews. Regarding the differences, the survey showed a higher urgency to autonomate alerts and user tips from instant messaging services and filtering the monitored outlets. The coordination with other fact-checkers and the analysis of the impact of the fact-checks was perceived as an issue with a lower urgency than it seemed from the interviews. Thus, these problems are considered specific to the region and need to be addressed locally.

\begin{figure}{
\centering
\includegraphics[width=0.75\linewidth]{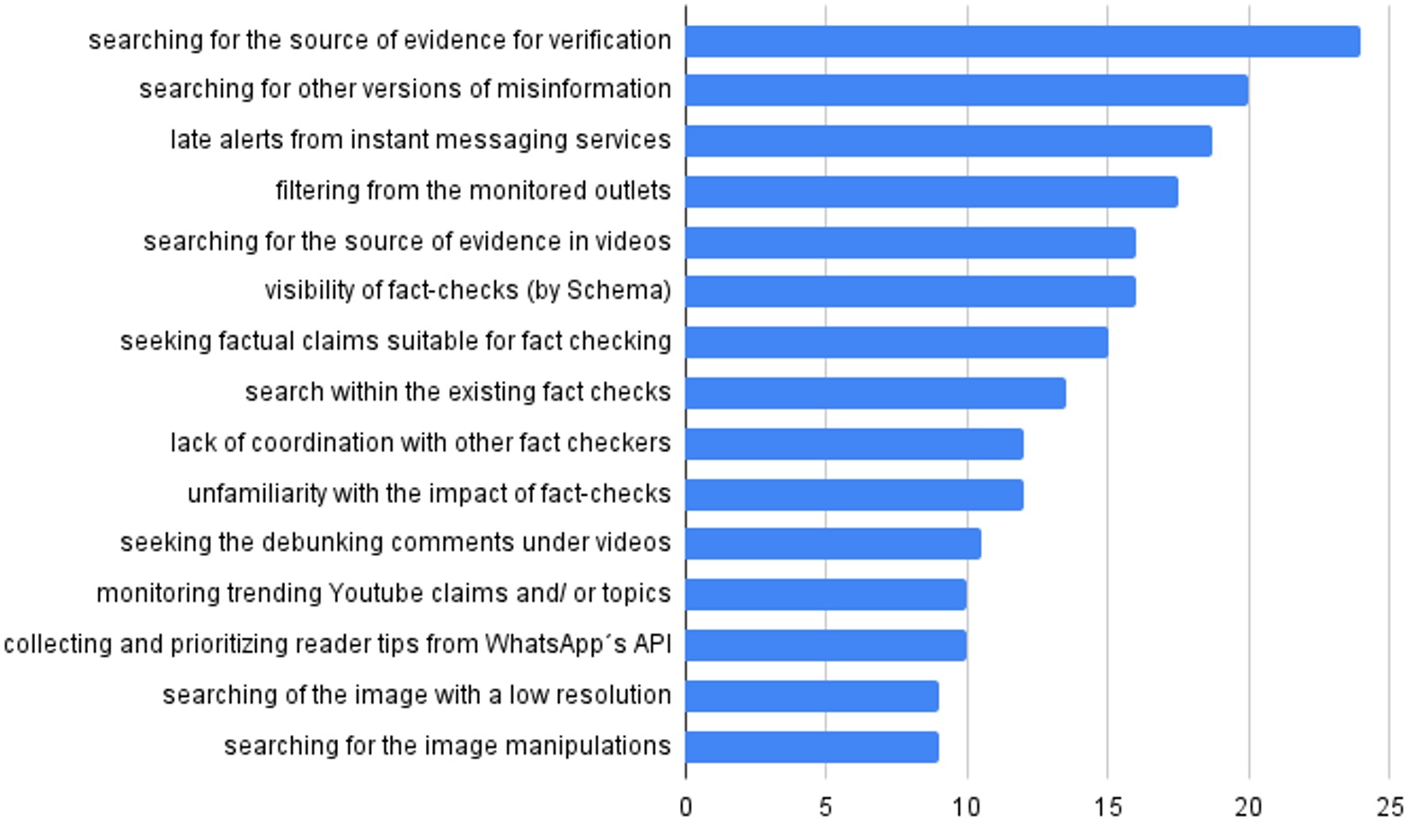}
\caption{The biggest problems of
European fact-checkers identified by the proposed coefficient of the perceived importance of the problem (scale of the coefficient ranges from 1 to 30)}
\label{fig:most-severe-pains}
\Description{The most severe challenge is searching for the source of evidence for verification}
}
\end{figure}

Addressing RQ2, our results (Fig.~\ref{fig:most-severe-pains}) indicate that the biggest problems of European fact-checkers lie in the monitoring as well as in the verification parts of the process. The interviewed fact-checkers are overloaded with potentially false content, which requires better filtering (beyond virality measures).

According to the results from interviews, the verification of the truthfullness of the claim is not perceived as an issue. However, some parts of the verification process take most of the valuable time of fact-checkers and would urgently need some AI support. The validation survey confirmed that one of the biggest problems relates to searching for the sources of evidence for verification, particularly in hard-to-find official documents, videos, and statistics. This is especially true for cases where data needs to be integrated, as one fact-checker (P10) noted:

\begin{quote}
\emph{``{[}I{]} hardly believe that machine recognition of
misinterpretation of scientific studies/statistics is possible.''}
\end{quote}

Even when misinformation is verified, its numerous versions exist and are shared on the internet. Therefore, the tools that would be able to identify other versions of the same misinformation would be crucial for misinformation mitigation. One fact-checker (P8) illustrated some parts of this process as follows:

\begin{quote}
\emph{``The content on Facebook is very hard to find. We have to use some tricks, but still, there is a lot of manual work. It is very hard to identify, given a text query, all the places and URL addresses where the content exists (external pages, Facebook posts\ldots{} images, and videos are even harder). No tool would do this and this consumes about half of our time\ldots. I think the technology to automate searching the source of misinformation already exist and with them, we would be able to do much more fact-checking.''}
\end{quote}

Additionally, two ideas were mentioned frequently by our respondents (10 times each):~

\begin{itemize}
\item
 a tool to monitor trending YouTube claims and/or topics, and~
\item
 a tool connected to WhatsApp's API which collects and prioritizes
 reader tips.
\end{itemize}

We further investigated the differences between the fact-checkers' problems according to the location, language group affiliation, and size
of their organization. 

\subsection{Differences between the fact-checkers' problems}

The fact-checking organizations that were respondents of our validation survey have been split into two groups: 12 Eastern (including Central) European organizations (from these countries: Czech Republic, Slovakia, Hungary,
Poland, Lithuania, Latvia, Estonia, Croatia, Slovenia), and 10 Western European organizations (from these countries: Belgium, Italy, Ireland, France, Spain, Germany, Switzerland, Austria, Denmark). We further distinguish 12 organizations from countries using low-resource languages (from Slavic and Ugro-Finnish language groups), and 10 organizations from countries with high-resource languages (from Germanic and Romance language groups). Finally, the size of organizations varied as follows: 3 large (250 and more employees), 3 medium (50 to 249 employees), 6 small (10 to 49 employees) and 9 micro-sized (less than 10 employees).

\subsubsection{Differences between the problems of Eastern and Western European fact-checkers}

\begin{figure}{
\centering
\includegraphics[width=0.85\linewidth]{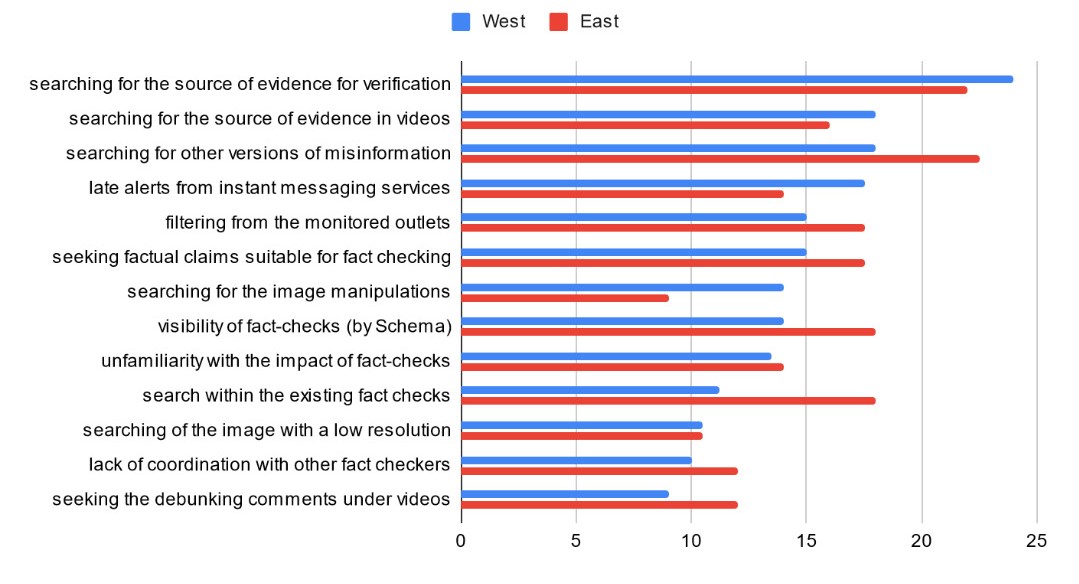}
\caption{Differences in the coefficients of the most important problems depending on the location of the fact-checkers (Eastern vs. Western Europe)}\label{fig:eastern-vs-western}
}
\Description{The biggest difference is in searching within the existing fact-checks}
\end{figure}

The data do not show many differences between the problems of Eastern and Western European fact-checkers (Fig.~\ref{fig:eastern-vs-western}). However, searching within the existing fact-checks (both their own and from other fact-checking organizations and in potentially different languages) is considered a more urgent problem to solve by Eastern European fact-checkers. This recognized need is relevant to address, as the existing fact-checks are an important source of evidence for these professionals. The
need for a tool that would help to autonomate this part of the process was confirmed by a fact-checker (P9) in a comment of the survey:

\begin{quote}
\emph{``Would totally be a great help. Compiling all the IFCN signatories' outputs into one big (keyword-driven) database? A fact-checker's dream. Although, it is fair to say that Google Search supplements a lot of this proposition -- although it's always difficult with other languages than English -- relevant now for Ukrainian.``}
\end{quote}

The first explanation of different perceptions of this difficulty is a better coverage of fact-checks in bigger language groups by Google Fact-Check Explorer. The lack of experience and/or resources to invest in ClaimReview markup was mentioned both by Eastern as well as some Western European fact-checkers.

The second explanation of this difference is that Eastern European fact-checkers need to check first the Western European fact-checks (which seems not to be the other way around). For example, a fact-checker (P7) revealed in the interview that:

\begin{quote}
~``\emph{Russian trolls know that in order to be successful, they have to go through the West (Germany, for example), because Poles do not like Russian sources\ldots{} Also, sometimes what is popular abroad -- mostly in Czech or English, such as vaccines -- is popular in Poland as well.''}
\end{quote}

In contrast, Western European fact-checkers focus more on misinformation modalities beyond text and platforms beyond Facebook.
This is demonstrated by their pressing need to autonomate alerts from instant messaging services, such as WhatsApp, possibly Telegram, and a more urgent need to search for image manipulations. The less perceived urgency of these needs by Eastern European respondents can be explained by the lack of capacity that allows them to focus mainly on text and Facebook, as mentioned during some of the interviews.

\subsubsection{Differences between the problems of European fact-checkers according to their language affiliation}

\begin{figure}{
\centering
\includegraphics[width=0.92\linewidth]{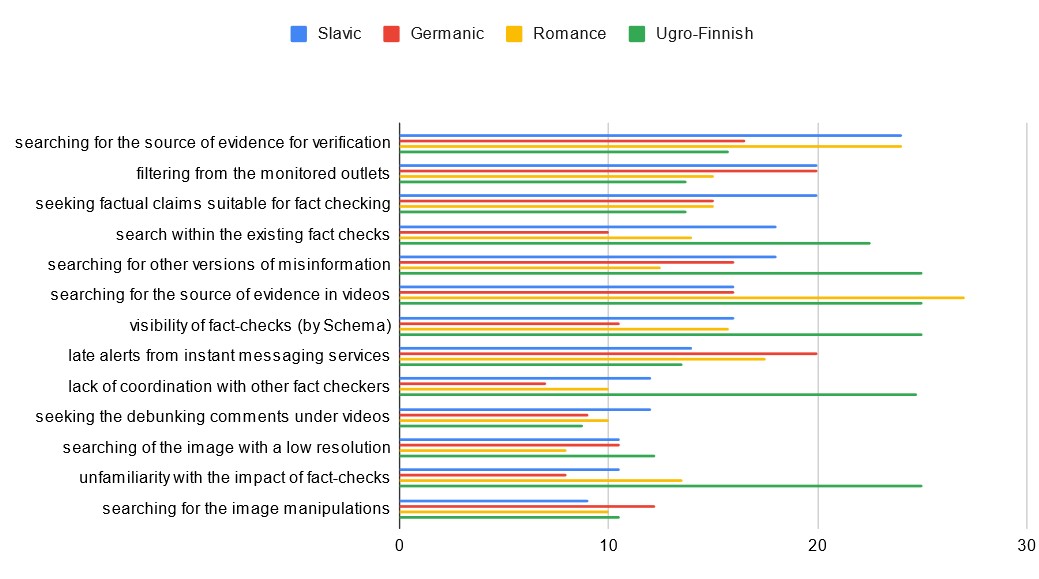}
\caption{Differences in the coefficients of the most important problems depending on fact-checkers' language group affiliation}\label{fig:language-groups}
}
\Description{The smallest language group perceived the highest urgency of the problems}
\end{figure}

Regarding the various language groups of the fact-checkers, we can see (Fig.~\ref{fig:language-groups}) that there are some differences in the perceived problems, but they are not always language-related. Nevertheless, the perceived urgency level stands out in the Ugro-Finnish group that is the smallest language group of participants. According to the results of the survey,
these professionals are much less involved in the coordination of actions of other fact-checkers. Generally, fact-checkers from the low resource languages, such as Hungarian, Slovak or Czech perceive more urgently the language-related problems in searching for the other versions of misinformation or in searching within the existing fact-checks. This is also in line with the NORDIS report~\cite{dierickx_report_2022}.

We can hypothesize that misinformation in dialects would be even harder to detect by machines than small language groups. This puts the people communicating in dialects in a more vulnerable position. A fact-checker (P10) supports this hypothesis:

\begin{quote}
\emph{``Machine does not recognise the dialect\ldots A tool which recognizes the dialect would help to simplify my work.''}
\end{quote}

\subsubsection{Differences between the problems of European fact-checkers according to the size of their organization}

\begin{figure}{
\centering
\includegraphics[width=0.8\linewidth]{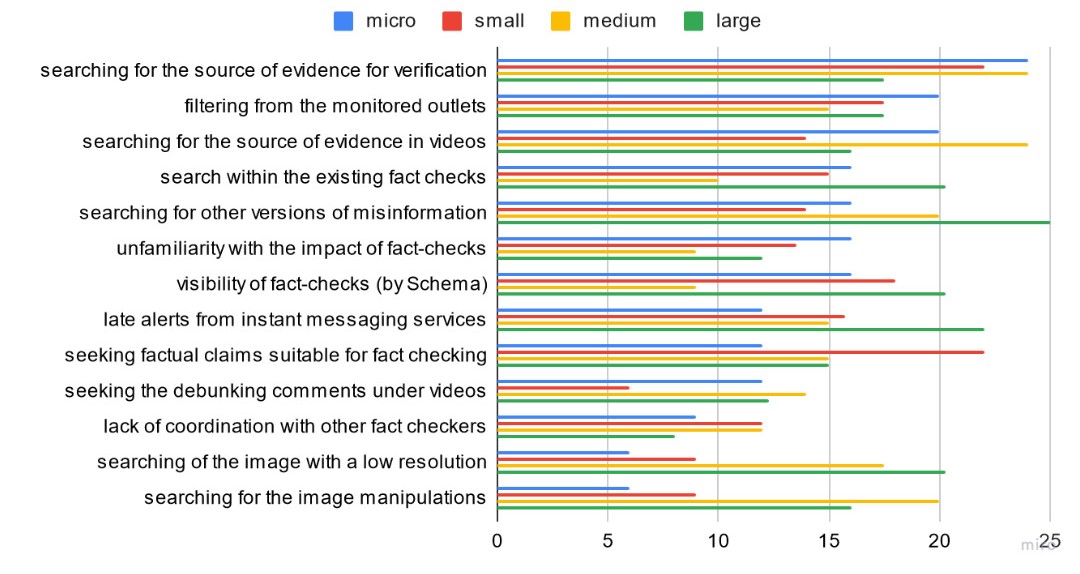}
\caption{Differences in the coefficients of the most important problems depending on fact-checking organization size}\label{fig:organization-size}
}
\Description{The urgency of the problems varies depending on the size of the organization}
\end{figure}

If we divide the results according to the size of the fact-checking organizations (Fig.~\ref{fig:organization-size}), we can see that large organizations perceive the tasks connected with: 1) image and video analysis; 2) monitoring of instant messaging services; as well as 3) searching for the other versions of misinformation as more difficult than smaller organizations do. The reason lies again in the higher capacities of these organizations to focus on more tasks that require more time, as confirmed during interviews. This finding points out that these tasks are not of little importance, but of little personal capacities to involve in such responsibilities. The autonomation of such tasks by artificial intelligence would help bigger organizations in the first place, but in the end also the smaller organizations, as the process of disinformation detection would be much easier and possible to complete with less capacity.

The other problems that were identified in our research include marketing issues (insights analysis, ClaimReview) that are just partially relevant for AI research. The reasons for these problems of both types of organizations are different: the small ones face capacity issues or lack of knowledge in terms of technology or marketing support; the fact-checkers of the largest organizations face complicated processes of the big media concerns that prevent fact-checkers to edit or monitor any content on the website that is common for all parts of the organization. Nevertheless, filling out ClaimReview would help AI researchers to
collect better datasets of the previous fact-checks and insights analysis would provide more information about the topics that are
important for users.

\section {Implications for AI research and AI-based tools}

\begin{figure}{
\centering
\includegraphics[width=1.0\linewidth]{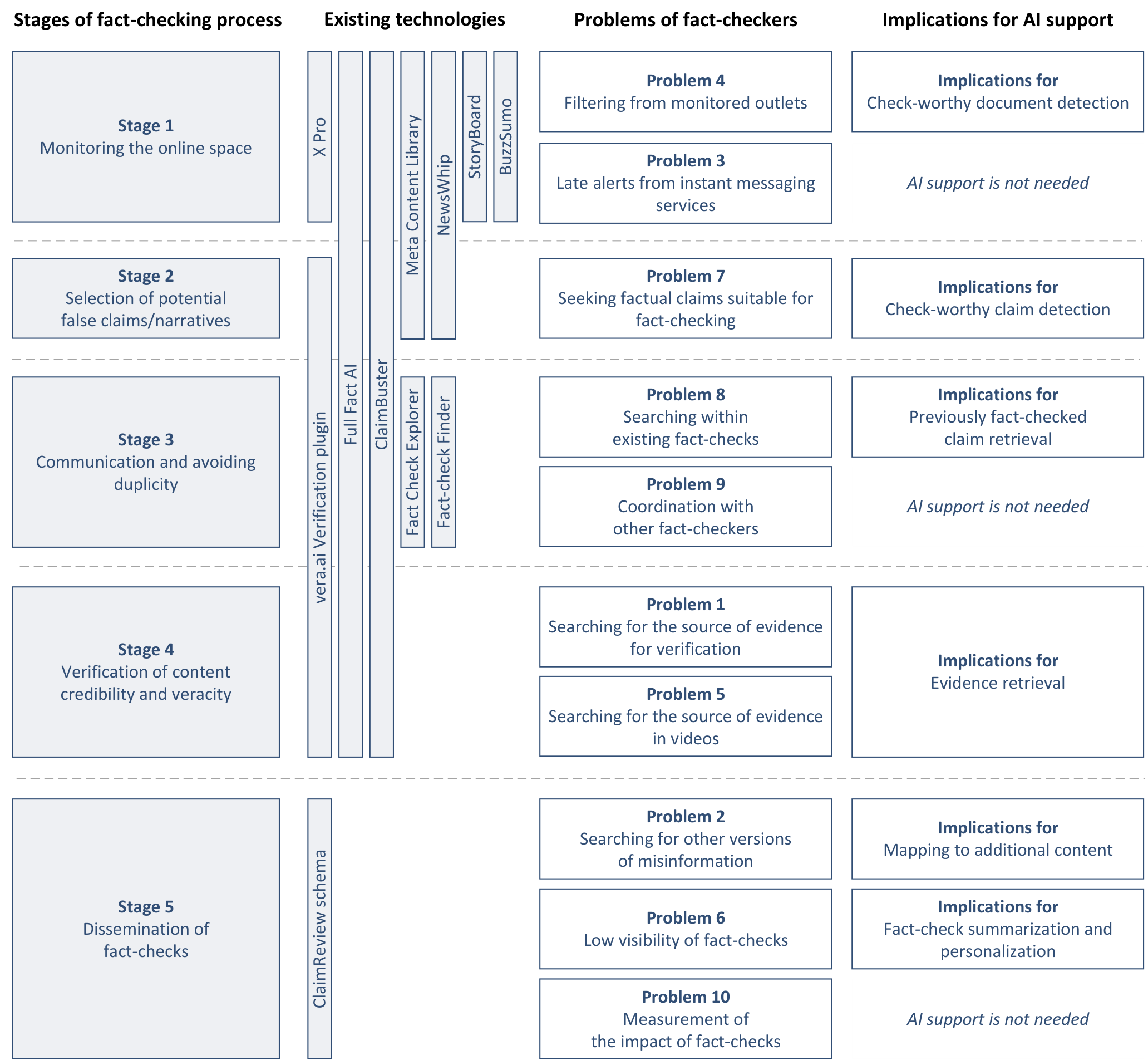}
\caption{Mapping of the top 10 most important problems of European fact-checkers with the corresponding stages of the fact-checking process, selected existing tools, and implications for AI support}\label{fig:mapping}
}
\Description{Stages of the fact-checking process and problems of fact-checkers mapped to the implications for AI tools}
\end{figure}

To answer RQ3, we identified implications and opportunities for research and development of AI tools that would support fact-checkers in fulfilling their tasks. For clarity, we mapped the stages of fact-checking process and the most urgent problems of fact-checkers (resulting from our validation survey) with the corresponding implications for AI support (Fig.~\ref{fig:mapping}). 

Our proposed implications for AI support come from the current state-of-the-art research and technical possibilities in the area of AI -- machine learning (ML) and NLP (including the latest development of generative large language models), and also from available datasets and tools and are consistent with the work by~\citet{nakov_automated_2021}. Compared to~\citet{nakov_automated_2021}, we add three more AI tasks: 1) \emph{check-worthy document detection}, 2) \emph{mapping of the existing fact-checks to additional (newly appearing) online content}, and 3) \emph{fact-check summarization and personalization}.

When discussing these implications, we aim to emphasize the need for human-centered AI systems (and a role that the current generative AI can play there) as well as reflect on the specific challenges and needs of the European fact-checkers identified in the previous sections. At the same time, we recognize that some of the most important problems do not require AI-based solutions. In such cases, the implementation of suitable tools represents mainly an engineering challenge (e.g., how to implement a tool that fact-checkers can use to insert ClaimReview schema into their fact-checks), which is out of the scope of this work.

\subsection{Implications for Check-worthy document detection}
The first AI task aims to support filtering the check-worthy documents (that is, news articles, blog posts, or even posts on social media) from monitored sources, which addresses the European fact-checkers' problem of filtering relevant content for fact-checking from the monitored outlets. It can be defined as follows: given an input set of documents, detect such documents that are worth fact-checking, and thus they should receive fact-checkers' attention (e.g., they contain factual claims, they are potentially impactful and harmful to society, etc.). This task can be addressed either as a classification (a document is or is not check-worthy) or a ranking problem (prioritizing the most check-worthy documents). It should be addressed with information available at document creation time (i.e., without relying on user feedback that appears later). Given that false information spreads faster than true information~\cite{vosoughi_spread_2018}, the impact of a fact-check can be limited once the post has already become viral. On the other hand, early fact-checks can limit the spread of false information and can thus serve as a way of pre-bunking, which has already been shown to be effective in existing studies~\cite{roozenbeek_psychological_2022}.

We argue that this task should generally precede the task of check-worthy claims detection (extraction) to limit and prioritize the amount of potential misinformation content the fact-checkers need to examine. Current systems (e.g., a fact-checking tool used within Meta's Third-Party Fact-Checking Program) often prioritize the virality of a post as noted by the fact-checkers in our as well as in the previous studies~\cite{noauthor_challenges_2020, dierickx_report_2022, micallef_true_2022}. In research work, check-worthiness detection is originally related to claims (i.e., identification of particular sentences, typically in political debates). Nevertheless, starting in 2020, the CheckThat! Lab introduced the detection of check-worthy tweets~\cite{hasanain_studying_2022}. Although tweets are natively short texts containing just a few sentences, we would like to emphasize the distinction with check-worthy claim detection (see below), and we consider this as the first step towards check-worthy document detection. At the moment, we are not aware of any works on check-worthiness detection for longer pieces of text (e.g., for social media posts from other platforms that do not pose such strict length limitations, or even for whole news articles/blogs) or for multimodal content. Such extension of current research works and the construction of a suitable dataset represents the possible next step.

For an AI solution to assist the fact-checkers, it needs to target sources relevant for them, consider criteria that are typically used by fact-checkers when they decide what to fact-check, and provide a means of justification/explanation of the selection of the check-worthy documents. The last two points can be facilitated by a related, more generic task of~\textit{credibility assessment}. Credibility signals (e.g., ones proposed by W3C Credible Web Community Group\footnote{\url{https://credweb.org/}}) would help fact-checkers (as well as other media professionals) to pre-screen the document and decide whether to proceed with the manual in-depth investigation to determine the necessity to fact-check it. Such a tool would speed up fact-checkers' comprehension of the online content. To detect such indicators, a wide variety of techniques may be used, from a simple lookup in whitelists/blacklists, through the automatic check of predefined credibility criteria (e.g., presence of an author's name, the known editorial board of a newspaper, etc.) up to advanced ML/NLP models. These would, for example, classify typographical and stylistic characteristics reflecting psychological features influencing the reader's sentiment, detect the use of logical fallacies, or classify the leaning (bias) of the text~\cite{martino_semeval-2020_2020, dhoju_differences_2019}. With the recent advancements of generative AI, namely large language models (LLMs), and their uptake for evaluation of various aspects of texts, such as text quality~\cite{hada-etal-2024-metal} or level of personalization~\cite{wang2023automated-evaluation, zugecova-etal-2025-evaluation}, the use of LLMs for credibility signals detection is also being researched, as discussed in a recent survey by~\citet{srba2024credsignals-survey}. One such example is the work of~\citet{leite2024llm-credsignals} which employed LLMs to detect 19 signals, such as the presence of bias, use of polarizing language, etc. The detected check-worthiness of documents can serve as one of the indicators if the detection also considers their potential harmfulness or their expected impact on society (see, e.g.,~\cite{alam_fighting_2021}), which are the aspects usually considered by the fact-checkers (see Fig.~\ref{fig:selection}).

\subsection{Implications for Check-worthy claims detection}

While the task of check-worthy document detection (and the related task of credibility assessment/credibility signals detection) has attracted researchers' attention only recently, the check-worthy claims detection has been addressed by the researchers for a longer period, and several established approaches, as well as datasets, already exist (see, e.g.,~\cite{hassan_toward_2017, srba2024credsignals-survey}). This task can be defined as follows: given an input sentence (or a set of multiple sentences, for example, from a political debate), detect such sentences that are worthy of fact-checking\footnote{While the definition of a check-worthy claim differs across literature, it is usually a factual statement, that the general public would be interested in knowing, whether it is true or not.}. Most of the existing approaches have so far focused on text using transformer-based language models, such as BERT~\cite{devlin_bert_2019}. However, in 2024 edition of the CLEF-2024 CheckThat! Lab task on check-worthy claims detection, the best-performing solutions were already using LLMs, including variations of LLaMA, Mistral, Mixtral, or GPT~\cite{checkthat2024}, which were either fine-tuned or used just in an in-context learning setting~\cite{checkthat2024-checkworthy}. There is also a growing need for multimodal approaches, since claims can be accompanied by images, they can themselves appear in images (e.g., a screenshot of a social media post), or images can modify/change their meanings. There are already the first multimodal datasets for claim detection (see, e.g.,~\cite{zlatkova_fact-checking_2019, Cheema_2022}).

Similar to the detection of check-worthy documents, for check-worthy claim detection solutions to be human-centered (and useful to the fact-checkers), they need to be able to justify why a given claim is considered check-worthy. This is where the current generative AI approaches could help by providing understandable justifications beyond simple highlighting of tokens/words considered important by the detector.

\subsection{Implications for Previously fact-checked claims retrieval}
The previously fact-checked claims retrieval has also attracted the researchers' attention lately; see e.g.,~\cite{hossain_covidlies_2020, shaar_overview_2021-1, nakov_overview_2022, mansour_did_2022, hardalov_crowdchecked_2022, pikuliak_multilingual_2023, ramponi2025multilingualvscrosslingualretrieval}. It is typically defined as a ranking problem, where, given an input claim and a set of verified claims (fact-checks), the goal is to retrieve a ranked list of previously fact-checked claims~\cite{shaar_that_2020}. 
However, the existing datasets and consequently, most of the existing approaches focus on retrieval in a single language -- mostly English -- although there are also datasets in other languages, such as the one by~\citet{kazemi_claim_2021}, MultiClaim~\cite{pikuliak_multilingual_2023}, or MMTweets~\cite{singh2024mmtweets}. 

Additionally, the existing datasets often miss cross-lingual links between fact-checks and social media posts. This limits their practical applicability since the claims can spread across borders and languages, which is especially relevant for the European fact-checkers. For example, \citet{pikuliak_multilingual_2023} acknowledge that there might be as many as seven times more pairs (many of those cross-lingual) in their dataset than what is actually identified. This is especially problematic given the results of our study, which suggest that searching for other versions of misinformation and searching in existing fact-checks is one of the most important problems among Central and Eastern European fact-checkers and fact-checkers from low-resource languages. The fact-checkers themselves could help foster this type of research by the more widespread use of ClaimReview and especially by providing more appearances of the fact-checked claim (using the \emph{ItemReviewed} field of the schema). However, this would require better supporting tools and/or training for fact-checkers as discussed in the sections above.

The role of LLMs in previously fact-checked claims retrieval is still relatively underexplored~\cite{vykopal2024llms-fact-checking}, although many solutions to the SemEval 2025 Task 7 task have already used an LLM in their pipeline~\cite{peng-etal-2025-semeval}. LLMs are currently typically used to rerank the retrieved verified claims~\cite{ramponi2025multilingualvscrosslingualretrieval}, to filter out irrelevant ones~\cite{vykopal2025largelanguagemodelsmultilingual}, or to decide if the input entails a given verified claim~\cite{choi2023automated-claim-matching}. Other options are to employ LLMs as a part of retrieval augmented generation system (RAG)~\cite{lewis2020rag}, or to use them indirectly, e.g., to normalize claims before retrieval~\cite{sundriyal-etal-2023-chaos} or to augment training datasets by synthetic data~\cite{singh2023utdrm}.

As part of this AI task, we developed a novel tool for fact-checkers -- Fact-check Finder\footnote{\url{https://fact-check-finder.kinit.sk/}} -- using a co-creation process in which identifying fact-checkers’ needs -- described in this study -- was the essential first step. We have collected so far the most extensive and the most linguistically diverse dataset of fact-checks to social media post pairs, experimented with and selected the best-performing text embedding models, and built the tool on top of them~\cite{pikuliak_multilingual_2023}. Fact-check Finder provides semantic cross-lingual search -- an important capability currently limited in Google Fact Check Explorer. For this purpose, Fact-check Finder performs content matching by means of calculating similarity between sentence embeddings created by multilingual language models~\cite{pikuliak_multilingual_2023}.

\subsection{Implications for Evidence retrieval}

Tools supporting evidence retrieval that would be explainable and trustworthy represent one of the most challenging but also one of the most urgent needs of European fact-checkers, as evidenced by the results of our study. Such tools would need to operate beyond Google searches on websites or extractions from Wikipedia. However, it is exactly these two that are often used when constructing datasets for claim verification; e.g., Wikipedia was used to construct the FEVER dataset and its variants~\cite{norregaard_danfever_2021, thorne_fever_2018, ullrich_csfever_2022}, and Google search was used to retrieve evidence in the case of the MultiFC~\cite{augenstein_multifc_2019} or X-Fact datasets~\cite{gupta_x-fact_2021}. 

Therefore, a more demanding data collection would be crucial for usable and credible verification tools for fact-checkers. One such attempt can be considered the dataset CTKFacts~\cite{ullrich_csfever_2022}, which extracted evidence from the articles of the Czech Press Agency (CTK), but databases containing (and combining) the most important sources for fact-checkers, such as official statistics, other official data, and the transcriptions of audio-visual content, are largely missing. Nevertheless, the introduction of OpenAI's automatic speech recognition system Whisper\footnote{\url{https://openai.com/research/whisper}} has a promising potential for multilingual audio-to-text transcriptions~\cite{radford_robust_2022}. 

There is also a potential for larger employment of LLMs, especially as a part of retrieval augmented systems (RAGs)~\cite{vykopal2024llms-fact-checking}, which prompts the LLMs to answer the input based on a set of retrieved examples and/or evidence instead of just relying on the LLMs' knowledge gained in the pre-training. For example,~\citet{momii-etal-2024-rag} employ RAG to retrieve evidence for an input claim by first generating questions that help narrow down the type of evidence needed to verify the input. The evidence can be retrieved from the whole web (as proposed, e.g., in~\cite{ullrich-etal-2024-aic}) as well as from specialized corpora (if available). On the other hand, relying purely on LLMs' pre-training to provide evidence is not advisable due to possible hallucinations~\cite{huang2025hallucinations} and the risks of using a conversational interface to attribute higher credibility to evidence provided in that way~\cite{anderl2024conversational}.

\subsection{Implications for Claim verification and justification production}
The claim verification and justification production, which require a human judgment, were not perceived as problematic by the European fact-checkers in our study. These professionals enjoyed such challenging tasks requiring higher cognitive load and creativity. Therefore, there is no urgency in developing tools that would directly aim to autonomate this stage and thus support or even replace the fact-checkers. Despite this fact, automatic claim verification and fact-checking is being actively researched, as also evidenced by an active community around the Fact Extraction and VERification Workshop (FEVER)~\cite{fever-2024-1}. We can see the community's shift towards the use of LLMs and RAGs for verdict production based on the retrieved evidence. While it may not be needed or even preferable due to ethical considerations to use the claim verification approaches directly, they can be used by the fact-checkers indirectly as supporting tools to rate or rerank the retrieved evidence and distinguish between sources that support or reject the claim under review~\cite{miranda_automated_2019}. Most recently,~\citet{warren2025-explainable-automated} examined the fact-checkers' requirements regarding the explanations provided by the automated fact-checking systems highlighting the gaps between the current state of the art and the practical needs of fact-checkers.

\subsection{Implications for Mapping to additional online content}
In the mapping to an additional online content task, the veracity, or knowledge of already fact-checked claims, is automatically disseminated to additional content (already existing or new, constantly emerging). This task is similar to the previously fact-checked retrieval, but, in this case, the input is an already fact-checked claim, and the output is a list of claim appearances, i.e., a list of social media posts or news articles in which the fact-checked claim is present and at the same time which has a positive stance (they support the claim; see~\cite{srba_monant_2022}). Alternatively, the task can be defined as a combination of previous fact-checked claim retrieval and stance classification. Such an additional step is not usually done in the manual fact-checking process, as it is impossible for the fact-checkers to manually find all existing relevant articles or routinely update their list as they continuously appear~\cite{wang_relevant_2018}. 

At the same time, searching for other versions of misinformation is the second most important problem as identified by the European fact-checkers surveyed in our study. Meta performs (or has performed until recently) this task to some extent -- when a new post shares an image that has been previously fact-checked, it propagates the original fact-check to that post. Having an autonomated method that would be able to work also with text in multilingual and cross-lingual settings would further help the fact-checkers increase the impact of their work but also help them monitor how the misinformation spreads and evolves. For this, more research on clustering the misinformation into broader narratives, their identification and evolution is needed~\cite{piskorski-etal-2025-semeval}.

\subsection{Implications for Fact-check summarization and personalization}

The recent emergence and consequent rapid development of generative AI, particularly large language models (LLMs), such as ChatGPT, provides an opportunity to support fact-check dissemination and tackle fact-checks' low visibility. Namely, LLMs excel in summarization~\cite{zhang2023benchmarking}, paraphrasing~\cite{tripto-etal-2024-ship} and translation~\cite{zhu2023multilingual} of the textual content. Therefore, we argue that LLMs can be employed to help fact-checkers process fact-checking articles into shorter and/or more comprehensive forms, such as fact-checking briefs that are suitable to be posted as a social media post. A step in this direction presents \textit{verdict production}, which employs LLMs to generate verdicts (explanations of why a claim is assessed as false or true) by summarizing existing fact-checks~\cite{russo-etal-2023-countering, russo2023-benchmarking}. These verdicts can then be employed on social media, e.g., as responses to social media posts spreading false claims.

Furthermore, we recognize the potential to prepare LLMs' prompts that will personalize the fact-checking article for the specific social media platform and target audience. In this way, fact-checkers can effectively generate a scaffold of the text that can be further manually adjusted (e.g., to clarify potentially incomplete information caused by automatic summarization). However, it is important to note that for such solutions to be human-centered, they need to respect and follow fact-checking organizations' guidelines on using LLMs and generative AI to create content.

\section {Discussion and study limitations}

In this study, we first studied the fact-checkers' activities, needs, and problems. We transformed the findings into practical implications for AI research, AI tasks, and AI-based tools. We have presented design implications, which have not yet been used to design and evaluate an AI tool. We defined the stages of the process in a natural setting, the most urgent user needs to autonomate, and rejections to do so. Obtaining such results was possible due to the interdisciplinary collaboration (information and computer scientists) as well as due to the involvement of users in the first stages of autonomation. The practical value of adopting a human-centric AI approach was exemplified by the Fact-check Finder (tool for previously fact-checked claim retrieval~\cite{pikuliak_multilingual_2023}). Our research study was the first stage of the HCAI design process, in which end-users are constantly involved in shaping and evaluating the supporting AI tools. Later on, fact-checkers should be actively involved in training AI models by applying a human-in-the-loop approach.

The implications in this study may be valuable not only for AI researchers but also for practitioners from technological companies and social media platforms. In general, we recognized that AI has a tremendous potential to help fact-checkers across the whole fact-checking process. Nevertheless, considering the current technical maturity, we suggest using its potential, especially in providing support in filtering the monitored content, searching within the existing fact-checks, and evidence retrieval. An interesting avenue of further research is its use for summarization of fact-checks and their personalization/adaptation for various platforms and/or users.

Providing multiplatform, multilingual, and multimodal solutions for fact-checkers would be the most useful help for this target group. Our findings extend the implications of~\citet{micallef_true_2022}, who propose personalized multiplatform solutions, including crowdsourcing of user tips, as well as of~\citet{nakov_automated_2021}, who discuss what the technology currently has to offer as well as current major challenges, such as multilingualism and multimodality. The results of our research also pointed to the scenarios that would be ideal but are currently not feasible to autonomate. First, filtering from monitored outlets based on the impact on society would be a task that requires long-term research on how misinformation affects society. Secondly, a tool that would be used globally by fact-checkers would be an ultimate goal for the multinational collaboration and exchange of misinformation narratives, but such a tool would be resource- and sustainability-heavy as well as context-specific. Thirdly, the ideal content verification would require the availability of (structured) data from governments, which is still not a reality. Fourthly, mapping to additional content to disseminate fact-checks would be impossible without the hands-on cooperation of fact-checkers with social media. 

We would also like to acknowledge limitations of our study. While we aimed to focus specifically on European fact-checkers and complement the existing works, we would like to encourage future researchers to investigate the processes and problems/needs of fact-checkers from the remaining underexplored regions in Asia and Africa. Covering these regions would be beneficial to get a comprehensive global picture of fact-checkers' routines and problems.

Although the research of the fact-checking activities and problems was directed to the European region, only minor differences were found when comparing these activities and problems with the findings about the fact-checkers from other non-English-speaking regions. The implications are also not in any way specific to individual countries and can be generalized to other contexts with similar attributes (e.g., countries where fact-checkers face the disinformation spread in other languages transmitting from the neighboring countries). This gives the implications a very broad applicability.

\section{Conclusions}

In this study, we analyze the activities, processes, and problems of fact-checkers. We focused on the European fact-checking space, and through this, covered many regions outside of the scope of the existing studies. We compared the details of our research with the findings of existing studies. We have unified the stages of the fact-checking process as it is understood in the current social-science and AI research. The hitherto known processes and common problems of this profession were summarized in conceptual models. The fact-checkers' needs inferred from problems served to inform future AI research on assisting fact-checkers. This is in accordance with the necessity to eliminate the recognized gap between current AI research and fact-checkers' requirements on and consequent utilization of AI-based tools.

We noted that monitoring the online space to find potential misinformation is one of the most time-consuming and difficult parts of the fact-checking process. As \textbf{this stage in the fact-checking process is often overlooked in AI research} (e.g.,~\cite{barron-cedeno_checkthat_2020, nakov_automated_2021, micallef_true_2022}, it is lacking the appropriate support from AI-based information technologies. Besides that, we postulate that without the AI-assisted additional step in the manual fact-checking process -- \textbf{mapping to additional online content} -- any effective mitigation of misinformation is impossible with such a scarce number of fact-checkers.

The biggest problems of European fact-checkers relate to \textbf{searching for the sources of evidence for verification}. We examined the information resources that the fact-checkers use, and our results show that the mass media are rarely used by fact-checkers and Wikipedia is never used and considered credible for fact-checking. Nevertheless, textual sources, such as news articles, academic papers, and Wikipedia documents, are one of the most commonly used types of evidence for automated fact-checking~\cite{guo_survey_2022}. Although Wikipedia may be useful for some tasks, \textbf{verification tools trained or tested just on Wikipedia and mass media datasets will never be sufficient for these professionals}. This calls both for the creation
of new datasets built on primary sources (official statistics, governmental reports, etc.) and for new methods and systems of supporting evidence retrieval that can tap into these hard-to-access resources.

Fact-checkers do not think that AI can automate their whole work, nor do they want it. Fact-checking often requires evaluating complex statements, talking to experts, or determining the truth, which are processes inherent to human judgment. Nonetheless,
fact-checkers are open and need assistance from AI-based tools, and in this paper, we summarized the possibilities that are offered by the current AI technologies.

Our findings confirm that the identified problems and activities of the target groups can and should serve as potential cases for autonomation to empower, not replace, people.

\section{Research ethical considerations}

All ideas and content presented in this research are the authors' own. This research involved human participants. Research planning, conduct, and reporting were consistent with local regulatory laws and regulations and aligned with ethical principles, such as the ACM Code of Ethics and Professional Conduct and international and national standards for such research. Participants were informed that their participation was voluntary and that they could withdraw at any time or request deletion of their data. They were briefed on the research objectives and the use of their data exclusively for research purposes. Additionally, participants were assured that their data would be anonymized, securely stored in the institutional repository, and deleted after publication. While approval from the Institutional Review Board/Ethics Committee was not obtained due to the absence of such a committee at our institution at the time of research design, every effort was made to ensure ethical conduct throughout the study.

\begin{acks}
This work was partially supported by CEDMO, a project funded by the European Union under the Contract No. 2020-EU-IA-0267; CEDMO 2.0, a project funded by the European Union under the Contract No. 101158609; AI-CODE, a project funded by the European Union under the Horizon Europe, GA No. 101135437; and by the EU NextGenerationEU through the Recovery and Resilience Plan for Slovakia under the project No. 09I01-03-V04-00006.
\end{acks}

\bibliographystyle{ACM-Reference-Format}
\bibliography{sources.bib}

\pagebreak
\appendix

\section{Semi-structured in-depth interview questions}\label{sec:annexes:interview-questions}

\textbf{Shortened introduction}

We would like to talk with you about the process of your fact-checking and about the problems it brings to you. Our aim is to identify how the appropriate AI support to fact-checkers can be provided.

\renewcommand{\arraystretch}{1.1}

\small
\begin{tabularx}{\linewidth}{ l X }
\caption{Semi-structured in-depth interview questions}\\\toprule\endfirsthead
\toprule\endhead
\midrule\multicolumn{2}{r}{\itshape continues on next page}\\\midrule\endfoot
\bottomrule\endlastfoot

\textbf{Category}                           & \textbf{Question}                                                                                 \\ \hline
\multirow{3}{8em}{Basic information}                  & Institution                                                                                    \\ \cline{2-2} 
                                   & Position                                                                                     \\ \cline{2-2} 
                                   & What kind of false information do you check?                                                                   \\ \hline
\multirow{5}{8em}{Problem and needs to autonomate}         & What do you miss the most when fact-checking?                                                                   \\ \cline{2-2} 
                                   & What is the most hard/ cumbersome when fact-checking?                                                               \\ \cline{2-2} 
                                   & What are the most repetitive works when fact-checking that could be automatized?                                                 \\ \cline{2-2} 
                                   & What are the most time consuming issues?                                                                     \\ \cline{2-2} 
                                   & What and how would you suggest improving the information systems that you use?                                                  \\ \hline
\multirow{4}{8em}{Monitoring and selection of potential misinformation} & How do you spot news that need to be checked?                                                                   \\ \cline{2-2} 
                                   & Do you also check the popularity of the claims before fact-checking? How do you check it? When is the right time for you to fact-check?                      \\ \cline{2-2} 
                                   & Where do you spot false information? Do you use any kind of resource management tools?                                              \\ \cline{2-2} 
                                   & Would it be beneficial to you if you had the (possibly most popular) check-worthy claims prepared by an AI system?                                \\ \hline
\multirow{7}{8em}{Verifying the content credibility and veracity}    & How do you verify whether the claim/ news are false or manipulated?                                                        \\ \cline{2-2} 
                                   & What kind of resources do you use to fact-check the news? Do you mention them in the fact-check?                                         \\ \cline{2-2} 
                                   & Which criteria do you use to verify the credibility of content?                                                          \\ \cline{2-2} 
                                   & Who verifies your fact-checks?                                                                          \\ \cline{2-2} 
                                   & What does your evaluation look like? Is it a scale or textual evaluation?                                                     \\ \cline{2-2} 
                                   & Would it be beneficial to you, if AI identifies some credibility criteria for you in the news (like e.g., missing author or sources, hateful sentiment, spell check errors etc.?) \\ \cline{2-2} 
                                   & What else would help you in verifying the content?                                                                \\ \hline
\multirow{7}{8em}{Communication and avoiding duplication}         & Which channels do you use for communication with other fact-checkers? Do you have a platform for communication?                                  \\ \cline{2-2} 
                                   & What type of communication is there? Do you also exchange some know-how there?                                                  \\ \cline{2-2} 
                                   & How do you organize your work between your colleagues?                                                              \\ \cline{2-2} 
                                   & Do you have any kind of system that you use in your organization to organize your workflow?                                            \\ \cline{2-2} 
                                   & How do you organize your work across fact-checkers in other organizations? Do you check whether the claim is already fact-checked?                        \\ \cline{2-2} 
                                   & Does it happen, there are duplications in fact-checking?                                                              \\ \cline{2-2} 
                                   & Would it be beneficial to you, if the information technology that you use checks whether the content is already fact-checked?                           \\ \hline
\multirow{5}{8em}{Dissemination of fact-checks}             & Do you publish your fact-checks just on your website or do you communicate them more widely?                                           \\ \cline{2-2} 
                                   & Do you also contact the person who made the misleading claim and ask them to correct or withdraw the claim?                                    \\ \cline{2-2} 
                                   & How is your fact-check structured? Do you also use semantics (like tags) to help search engines identify the claims? Which tools do you use to make the semantic markup?     \\ \cline{2-2} 
                                   & Do you cooperate with social media? How? Which languages do you cover?                                                      \\ \cline{2-2} 
                                   & What would be helpful in dissemination of your fact-checks?                                                            \\ \hline
Other                                 & \textit{Unstructured discussion}
\end{tabularx}

\section{Quantitative validation survey questions}\label{sec:annexes:survey-questions}

\textbf{Introduction}

Dear fact-checker, dear editor,

We have collected the most serious problems that were mentioned during our
interviews with Central European fact-checkers, operating in Slavic
languages. This survey is meant to collect the answers of the
fact-checkers of the rest of Europe to become a more complex picture of
the needs and problems of fact-checkers. As we plan to publish the results
as a research paper, your answers may serve as important inputs for the
AI research community to research and develop better solutions for you
and to help your processes be smarter and smoother. Therefore, please,
indicate the level and frequency of problem felt during your fact-checking
process, as well as the perceived priority of support needed by a tool /
tech. assistant for your work tasks that take you the most of the time
or are most repetitive. Thank you very much for your valuable answers as
well as for your important work.

\renewcommand{\arraystretch}{1.1}

\small
\begin{tabularx}{\linewidth}{ l X X }
\caption{Quantitative validation survey questions}\\\toprule\endfirsthead
\toprule\endhead
\midrule\multicolumn{3}{r}{\itshape continues on next page}\\\midrule\endfoot
\bottomrule\endlastfoot

  \textbf{\#}       & \textbf{Question}                                                                                                                                      & Answer options                                                                                                                                                      \\ \midrule
I. & Institution & Free text answer \\ \midrule

II. & How big is your institution? & \begin{tabular}[c]{@{}l@{}}1) micro (fewer than 10 employees)\\ 2) small (10 to 49 employees)\\ 3) medium-sized (50 to 249 employees)\\ 4) large (250 and more employees)\end{tabular} \\ \midrule

\multirow{2}{*}{III.} & How often do you need to filter from monitored outlets manually to decide what to focus on? \par \vspace{0.5em} \textit{Example: You are overloaded with potential disinformation from e.g., Crowd Tangle. You need to filter them to see just the results „this looks suspicious“, „this might be important“...)} \vspace{2pt} & \multirow{2}{*}{\begin{tabular}[c]{@{}l@{}}1) More times a day\\ 2) About once a day\\ 3) About once a week\\ 4) About once a month\\ 5) Less than once a month\\ 6) Never\end{tabular}} \\ & \\ \midrule

\multirow{2}{*}{IV.} & If it was the case, how seriously do you suffer from manual filtering from monitored outlets? Would you appreciate some tech. support (tools) for this work? & 5 point Likert scale: 1 = No problem, I like to do it; 5 = I perceive it as a big problem. I really need a help with this \\ \cmidrule(l){2-3} & Please, provide us comments, if you have any & Free text answer \\ \midrule

\multirow{2}{*}{V.} & How often would you need to have any alerts about potential disinformation from instant messaging services?  \par \vspace{0.5em} \textit{Example: You noticed that there are many screenshots from Telegram that are shared on Facebook and you would like to have a quicker alert before it is shared heavily on Facebook}  \vspace{2pt} & \multirow{2}{*}{\begin{tabular}[c]{@{}l@{}}1) More times a day\\ 2) About once a day\\ 3) About once a week\\ 4) About once a month\\ 5) Less than once a month\\ 6) Never\end{tabular}} \\ & \\ \midrule

\multirow{2}{*}{VI.}  & If it was the case, how seriously do you suffer from the late alerts about potential disinformation from instant messaging services? Would you appreciate some tech. support (tools) for this work? & 5 point Likert scale: 1 = No problem, I like to do it; 5 = I perceive it as a big problem. I really need a help with this \\ \cmidrule(l){2-3} & Please, provide us comments, if you have any & Free text answer \\ \midrule

\multirow{2}{*}{VII.} & How often do you need to seek factual claims suitable for fact-checking in a selected article? \par \vspace{0.5em} \textit{Example: You are overloaded by potential disinformation and you need to filter out just the factual claims that need/can be fact-checked} \vspace{2pt} & \multirow{2}{*}{\begin{tabular}[c]{@{}l@{}}1) More times a day\\ 2) About once a day\\ 3) About once a week\\ 4) About once a month\\ 5) Less than once a month\\ 6) Never\end{tabular}} \\ & \\ \midrule

\multirow{2}{*}{VIII.} & If it was the case, how seriously do you suffer from seeking factual claims suitable for fact-checking in a selected article? Would you appreciate some tech. support (tools) for this work? & 5 point Likert scale: 1 = No problem, I like to do it; 5 = I perceive it as a big problem. I really need a help with this \\ \cmidrule(l){2-3} & Please, provide us comments, if you have any & Free text answer \\ \midrule

\multirow{2}{*}{IX.} & How often would you need to coordinate with other fact-checking organizations to avoid duplicates of your work? \par \vspace{0.5em} \textit{Example: You want to be aware, who is doing what, not to do the duplicate fact-checks. Better coordination with the other fact-checking organizations is needed} \vspace{2pt} & \multirow{2}{*}{\begin{tabular}[c]{@{}l@{}}1) More times a day\\ 2) About once a day\\ 3) About once a week\\ 4) About once a month\\ 5) Less than once a month\\ 6) Never\end{tabular}} \\ & \\ \midrule

\multirow{2}{*}{X.} & If it was the case, how seriously do you suffer from duplicates of your work with other fact-checking organizations? Would you appreciate some tech. support for better coordination with the other fact-checking organizations? & 5 point Likert scale: 1 = No problem, I like to do it; 5 = I perceive it as a big problem. I really need a help with this \\ \cmidrule(l){2-3} & Please, provide us comments, if you have any & Free text answer \\ \midrule

\multirow{2}{*}{XI.} & How often do you need to search for the source of evidence for verification of the potential disinformation? \par \vspace{0.5em} \textit{Example: You need to find the relevant proof that the information you are fact-checking is manipulated, not true etc. You need a better search in the official statistics, media etc...to fulfill this task.} & \multirow{2}{*}{\begin{tabular}[c]{@{}l@{}}1) More times a day\\ 2) About once a day\\ 3) About once a week\\ 4) About once a month\\ 5) Less than once a month\\ 6) Never\end{tabular}} \\ & \\ \midrule

\multirow{2}{*}{XII.} & If it was the case, how seriously do you suffer from searching for the source of evidence for verification of the potential disinformation? Would you appreciate some tech. support (tools) for this work? & 5 point Likert scale: 1 = No problem, I like to do it; 5 = I perceive it as a big problem. I really need a help with this \\ \cmidrule(l){2-3} & Please, provide us comments, if you have any & Free text answer \\ \midrule

\multirow{2}{*}{XIII.} & How often do you need to search for the source of evidence in videos? \par \vspace{0.5em} \textit{Example: You need to verify a very toxic rumor about a politician and you know, you will find the proof in parliamentary speeches. But it is very hard to search within these materials (videos without appropriate metadata). You would need a searchable textual transcript of the video.} & 1) More times a day \par 2) About once a day \par 3) About once a week \par 4) About once a month \par 5) Less than once a month \par 6) Never \\ \midrule

\multirow{2}{*}{XIV.} & If it was the case, how seriously do you suffer from searching for the source of evidence in videos? Would you appreciate some tech. support (tools) for this work? & 5 point Likert scale: 1 = No problem, I like to do it; 5 = I perceive it as a big problem. I really need a help with this \\ \cmidrule(l){2-3} & Please, provide us comments, if you have any & Free text answer \\ \midrule

\multirow{2}{*}{XV.} & How often do you need to seek the debunking comments of common users under videos? \par \vspace{0.5em} \textit{Example: Some comments under manipulative Youtube videos (e.g., with links) debunk false information that the video contains. You need to quickly find the potential debunking comments to be able to assess the videos faster. } & \multirow{2}{*}{\begin{tabular}[c]{@{}l@{}}1) More times a day\\ 2) About once a day\\ 3) About once a week\\ 4) About once a month\\ 5) Less than once a month\\ 6) Never\end{tabular}} \\ & \\ \midrule

\multirow{2}{*}{XVI.} & If it was the case, how seriously do you suffer from seeking the debunking comments of common users under videos? Would you appreciate some tech. support (tools) for this work?                                                      & 5 point Likert scale: 1 = No problem, I like to do it; 5 = I perceive it as a big problem. I really need a help with this \\ \cmidrule(l){2-3} & Please, provide us comments, if you have any & Free text answer \\ \midrule

\multirow{2}{*}{XVII.} & How often do you need to reverse-search the image with a low resolution? \par \vspace{0.5em} \textit{Example: A lot of disinformation in your country is shared on images with low resolution that Google reverse-search doesn't capture. You need a textual explanation, what is there to be able to search within them.} & \multirow{2}{*}{\begin{tabular}[c]{@{}l@{}}1) More times a day\\ 2) About once a day\\ 3) About once a week\\ 4) About once a month\\ 5) Less than once a month\\ 6) Never\end{tabular}} \\
& \\ \midrule
             
\multirow{2}{*}{XVIII.} & If it was the case, how seriously do you suffer from searching for an image with a low resolution? Would you appreciate some tech. support (tools) for this work? & 5 point Likert scale: 1 = No problem, I like to do it; 5 = I perceive it as a big problem. I really need a help with this \\ \cmidrule(l){2-3} & Please, provide us comments, if you have any & Free text answer \\ \midrule

\multirow{2}{*}{XIX.}  & How often do you need to search for the image manipulations within the images? \par \vspace{0.5em} \textit{Example: A lot of filters on manipulated images were identified by a tool for filter detection. But many filters are just used for aesthetic reasons. You need to find the ones that are relevant for more serious manipulations detections} \vspace{2pt} & \multirow{2}{*}{\begin{tabular}[c]{@{}l@{}}1) More times a day\\ 2) About once a day\\ 3) About once a week\\ 4) About once a month\\ 5) Less than once a month\\ 6) Never\end{tabular}} \\ \midrule 

\multirow{2}{*}{XX.} & If it was the case, how seriously do you suffer from searching for the image manipulations within the images? Would you appreciate some tech. support (tools) for this work? & 5 point Likert scale: 1 = No problem, I like to do it; 5 = I perceive it as a big problem. I really need a help with this \\ \cmidrule(l){2-3} & Please, provide us comments, if you have any & Free text answer \\ \midrule

\multirow{2}{*}{XXI.} & How often do you need to search within the existing fact-checks (your own and/ or of the other organizations)? \par \vspace{0.5em} \textit{Example: The politician, whose speech you are fact-checking, made the same misleading claim as a month ago. You need to search in your previous fact-checks to find the same sources of evidence.} & \multirow{2}{*}{\begin{tabular}[c]{@{}l@{}}1) More times a day\\ 2) About once a day\\ 3) About once a week\\ 4) About once a month\\ 5) Less than once a month\\ 6) Never\end{tabular}} \\ & \\ \midrule

\multirow{2}{*}{XXII.} & If it was the case, how seriously do you suffer from searching within the existing fact-checks (your own and/ or of the other organizations)? Would you appreciate some tech. support (tools) for this work? & 5 point Likert scale: 1 = No problem, I like to do it; 5 = I perceive it as a big problem. I really need a help with this \\ \cmidrule(l){2-3} & Please, provide us comments, if you have any & Free text answer \\ \midrule

\multirow{2}{*}{XXIII.} & How often do you need to search for the other versions of the same disinformation to debunk them all/ to have a better outreach? \par \vspace{0.5em} \textit{Example: You have debunked a disinformation, but you are aware that it is shared in slightly different forms on other websites or social media. You want to find as much similar disinformation as possible to connect it with your debunk.} & \multirow{2}{*}{\begin{tabular}[c]{@{}l@{}}1) More times a day\\ 2) About once a day\\ 3) About once a week\\ 4) About once a month\\ 5) Less than once a month\\ 6) Never\end{tabular}} \\ & \\ \midrule

\multirow{2}{*}{XXIV.} & If it was the case, how seriously do you suffer from searching for the other versions of the same disinformation to debunk them all/ to have a better outreach? Would you appreciate some tech. support (tools) for your work? & 5 point Likert scale: 1 = No problem, I like to do it; 5 = I perceive it as a big problem. I really need a help with this \\ \cmidrule(l){2-3} & Please, provide us comments, if you have any & Free text answer \\ \midrule

\multirow{3}{*}{XXV.} & How often do you analyze the insights/ impact of your fact-checks? \par \vspace{0.5em} \textit{Example 1: You need to learn which of your debunks do your users read.} \par \vspace{0.5em} \textit{Example 2: You wish to seek duplicates of your fact-checks (post hoc)} & \multirow{2}{*}{\begin{tabular}[c]{@{}l@{}}1) More times a day\\ 2) About once a day\\ 3) About once a week\\ 4) About once a month\\ 5) Less than once a month\\ 6) Never\end{tabular}} \\ & \\ \midrule

\multirow{2}{*}{XXVI.} & If you don't analyze, how seriously do you suffer from the lack of information about the impact of your fact-checks? Would you appreciate some help with the assessment of the impact of your fact-checks? & 5 point Likert scale: 1 = No problem, we don't need it; 5 = I perceive it as a big problem. I really need a help with this \\ \cmidrule(l){2-3} & Please, provide us comments, if you have any & Free text answer \\ \midrule

\multirow{2}{*}{XXVII.} & Do you wish to make your fact-checks more visible in Google search results or in Google Fact Check Explorer? \par \vspace{0.5em} \textit{Example: You would like to use the structure in ClaimReview Schema on your website, but you don't have any technical capacities or possibilities to do it (even in the Wordpress plugin from Fullfact -- ClaimReview Schema)} & \multirow{2}{*}{\begin{tabular}[c]{@{}l@{}}1) More times a day\\ 2) About once a day\\ 3) About once a week\\ 4) About once a month\\ 5) Less than once a month\\ 6) Never\end{tabular}} \\ & \\ \midrule

\multirow{2}{*}{XXVIII.} & Would you appreciate some tech. support to make your fact-checks more visible in Google search results or in Google Fact Check Explorer? & 5 point Likert scale:1 = No, I want to do it on my own; 5 = I perceive it as a big problem. I really need a help with this \\ \cmidrule(l){2-3} & Please, provide us comments, if you have any & Free text answer \\ \midrule

XXIX. & Do you have any ideas about other tools that would help to simplify your work? Please don’t feel limited by these examples and fill in all your ideas. & Multiple choice answer: \par 1) A tool to monitor trending Youtube claims and/or topics \par 2) A tool connected to WhatsApp's API which collects and prioritizes reader tips \par 3) No, the above-mentioned needs covered everything that I can think of \par 4) Other (free text answer) \\ \midrule

XXX. & Are you interested in receiving the results of this research? & 1) yes, as a preprint \par 2) yes, as a final paper \par 3) no, thanks
\end{tabularx}

\end{document}